\documentclass[reprint,amsmath,amssymb,aps,longbibliography]{revtex4-2}
\usepackage{graphicx}
\graphicspath{{images/}}
\usepackage{float}
\usepackage{adjustbox}
\usepackage{verbatim}

\usepackage{upgreek}
\usepackage{graphicx}
\usepackage{dcolumn}
\usepackage{bm}
\usepackage{color}
\usepackage{mathtools}
\usepackage[normalem]{ulem}
\usepackage[english]{babel}
\usepackage{xr-hyper}
\usepackage[colorlinks=true,linkcolor=blue,urlcolor=blue,citecolor=blue]{hyperref}

\newcommand{\del}[1]{\sout{#1}}  
\renewcommand{\del}[1]{}  

\newcommand{\fig}[1]{Fig.~\ref{#1}}
\newcommand{\figs}[2]{Figs.~\ref{#1} and \ref{#2}}
\newcommand{\Fig}[1]{Figure~\ref{#1}}

\newcommand{\eqs}[2]{Eqs.~(\ref{#1}) and (\ref{#2})}

\renewcommand{\section}[1]{{\bf{#1:}}}

\begin{document}
	
	\title{Diffusion coefficient power laws and defect-driven glassy dynamics in swap acceleration}
	
	\author{Gautham Gopinath$^1$}
	\thanks{Present address: Department of Physics, Yale University, New Haven, Connecticut 06511, USA.}
	\author{Chun-Shing Lee$^1$}
	\author{Xin-Yuan Gao$^1$}
	\author{Xiao-Dong An$^1$}
	\author{Chor-Hoi Chan$^2$}
	\author{Cho-Tung Yip$^2$}
	\email[Email: ]{h0260416@hit.edu.cn}
	\author{Hai-Yao Deng$^3$}
	\email[Email: ]{dengh4@cardiff.ac.uk}
	\author{Chi-Hang Lam$^1$}
	\email[Email: ]{C.H.Lam@polyu.edu.hk}
	
	\address{
		$^1$Department of Applied Physics, Hong Kong Polytechnic University, Hung Hom, Hong Kong, China \\
		$^2$Department of Physics, Harbin Institute of Technology, Shenzhen 518055, China\\
		$^3$School of Physics and Astronomy, Cardiff University, 5 The Parade, Cardiff CF24 3AA, Wales, UK}

	
	\begin{abstract}
      Particle swaps can drastically accelerate dynamics in glass. The mechanism is expected to be vital for a fundamental understanding of glassy dynamics.
To extract defining features, we propose a partial swap model with a fraction $\phi_s$ of swap-initiating particles, which can only swap locally with each other or with regular particles.  We focus on the swap-dominating regime.
At all temperatures studied, particle diffusion coefficients scale with $\phi_s$ in unexpected power laws with temperature-dependent exponents, consistent with the kinetic picture of glassy dynamics. At small $\phi_s$, swap-initiators, becoming defect particles, induce remarkably typical glassy dynamics of regular particles. This supports defect models of glass.

\end{abstract}
	
	\maketitle 
	
	Under rapid cooling, most liquids experience a considerable dynamic slowdown accompanied by an increase in viscosity \cite{biroli2013review,arceri2020}.
Despite decades of research and significant progress, an encompassing theoretical description of this dynamic arrest has proven elusive.     Even the most fundamental questions, such as whether the arrest is of thermodynamic or kinetic origin, are still under heated debate \cite{stillinger2013review}. 
Complementary to experimental approaches, 
molecular dynamics (MD) simulations play a pivotal role because they enable one to examine the microscopic dynamics directly.
For a long time, preparation of equilibrium MD systems in the deeply supercooled regime was challenging due to the slow dynamics.
Recently, lots of attention has been focused on a swap Monte-Carlo algorithm \cite{grigera2001,procaccia2015,berthier2016,turci2017}, which speeds up the equilibration of polydisperse fluids by over ten orders of magnitude ~\cite{ninarello2017}.
A further ingenious observation by Wyart and Cates  \cite{wyart2017} is that the extraordinary speedup is not only useful technically, but has strong theoretical implications on the underlying mechanism of glassy dynamics. They argue that the success of swap evidences the kinetic school  against the thermodynamic picture. This has initiated an interesting debate \cite{berthier2019theory}. The advancements on swap acceleration have spurred a flurry of theoretical works 
on possible explanations based on a variety of approaches \cite{szamel2018,brito2018,ikeda2017,gutierrez2019}.
Existing simulation results, aiming predominantly at computational efficiency, lack hallmark features which can conclusively discriminate the theories and settle the debate.


In this work, we study generalizations of the swap algorithm which are not necessarily the most efficient, aiming rather at a better understanding of both swap and glassy dynamics. 
We consider polydisperse soft repulsive particles in two dimensions. Hybrid dynamics of MD evolution with periodic Monte Carlo swap attempts is applied. We adopt particle swap following Refs.~\cite{grigera2001,procaccia2015}, rather than radius swap \cite{ninarello2017} so that all particle attributes including positions are swapped. Swaps thus contribute to particle movements directly. In addition, we consider local, i.e. nearest neighboring, swaps by restricting the swaps to particle pairs within a short distance of the order of particle diameters. Local and non-local swaps have been found to generate similar results \cite{ninarello2017}. This local particle swapping scheme, without long jumps, leads to realistic like particle dynamics in which standard measures such as the diffusion coefficient are well defined.

\section{Partial swap}
A key feature of our model is that only certain particle pairs enjoy swap attempts. 
Before a simulation starts, we randomly select a fraction $\phi_s$ of particles in the system, referred to as {\it swap-initiators}. Only these particles can perform swaps with themselves or with other regular particles. Regular particles cannot swap directly among themselves.
Previous studies are thus akin to $\phi_s=1$ \cite{grigera2001,procaccia2015,berthier2016,turci2017,ninarello2017}.
By varying $\phi_s$, an expanded parameter space is explored. At small $\phi_s$, one gets a system of regular particles with a small density of swap-initiators as defects.
In our study, we focus on a swap-dominating regime by applying a sufficiently high swap attempt frequency,
so that the diffusion coefficient of the regular particles must be enhanced by swaps by  at least 10 times. 
MD steps then only lead to negligible direct motions but are essential to position the particles for effective swaps.

We use dimensionless units in which the average particle diameter, the particle mass and the Boltzmann constant are all set to 1.
Particle diameters follow a uniform distribution with a standard deviation of 0.18 to ensure disordered arrangements. Two particles separated by a distance $r$ interact with a repulsive pair potential ${\sim}r^{-12}$. Unit particle density is considered. 
Our local swaps can only exchange the positions of particles within a cutoff distance of 1.5, the first minimum in the particle pair distribution function. Swaps are conducted using standard Monte Carlo algorithms following detailed balance so that the system thermodynamics is exactly preserved while dynamics is dramatically accelerated.
Further details 
can be found in supplementary information (SI) \cite{supplement, fernandez2007, berthier2019, thompson2021lammps, sadigh2012, ovito, narumi2011, shi2020, lulli2021, lee2021, gao2022}.


	\begin{figure}[tb]
		\includegraphics[width=0.8\columnwidth]{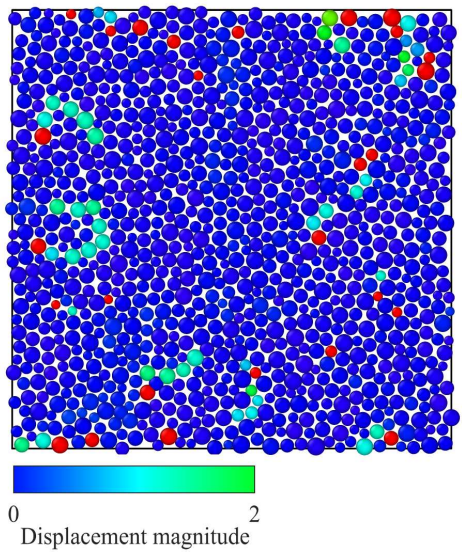}
		\caption{A system with a fraction $\phi_s=0.026$ of swap-initiating particles (red circles), only which can exchange positions with regular particles or with each other. The regular particles are color coded according to their displacements over a duration 0.015 at $T=0.16$. String-like motions can be observed. 
        }
		\label{fig:system}
	\end{figure}

\section{Results}
    Figure \ref{fig:system} shows an example of the overall system consisting of a small density of swap-initiators (red) in a background of regular particles (blue to green). It is seen that regular particles with larger displacements (light blue to green) are those close to the swap-initiators, as the latter induce these motions. 
    We measure the diffusion coefficient of the regular particles  defined as
		$D_r = {\langle |\textbf{r}_i(t)-\textbf{r}_i(0)|^2 \rangle}/{4t}$
measured at long time $t$, where $\textbf{r}_i(t)$ is the position of regular particle $i$.
	\begin{figure}[t]
		\centering
			\includegraphics[width=\columnwidth]{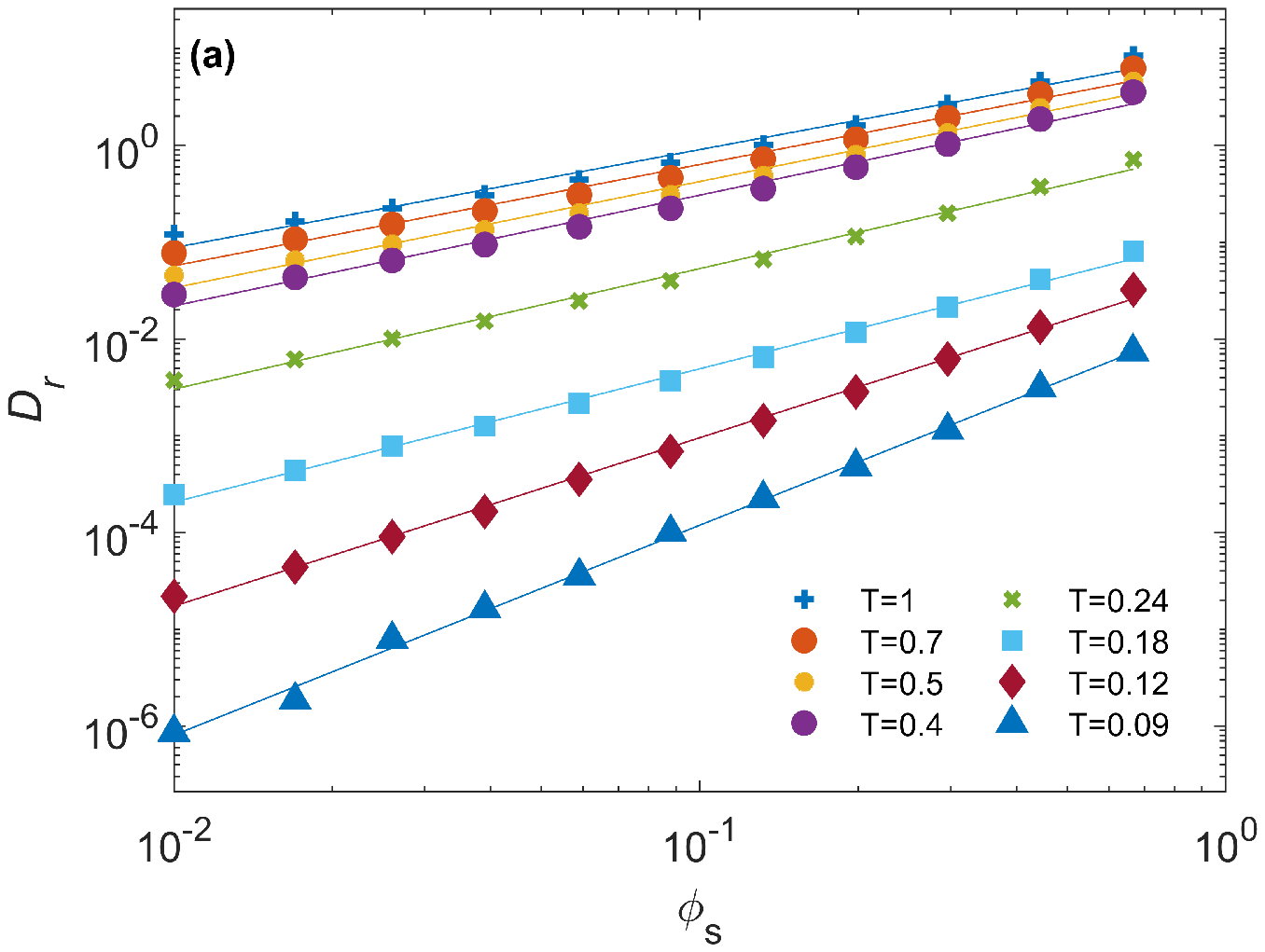}	
			\includegraphics[width=\columnwidth]{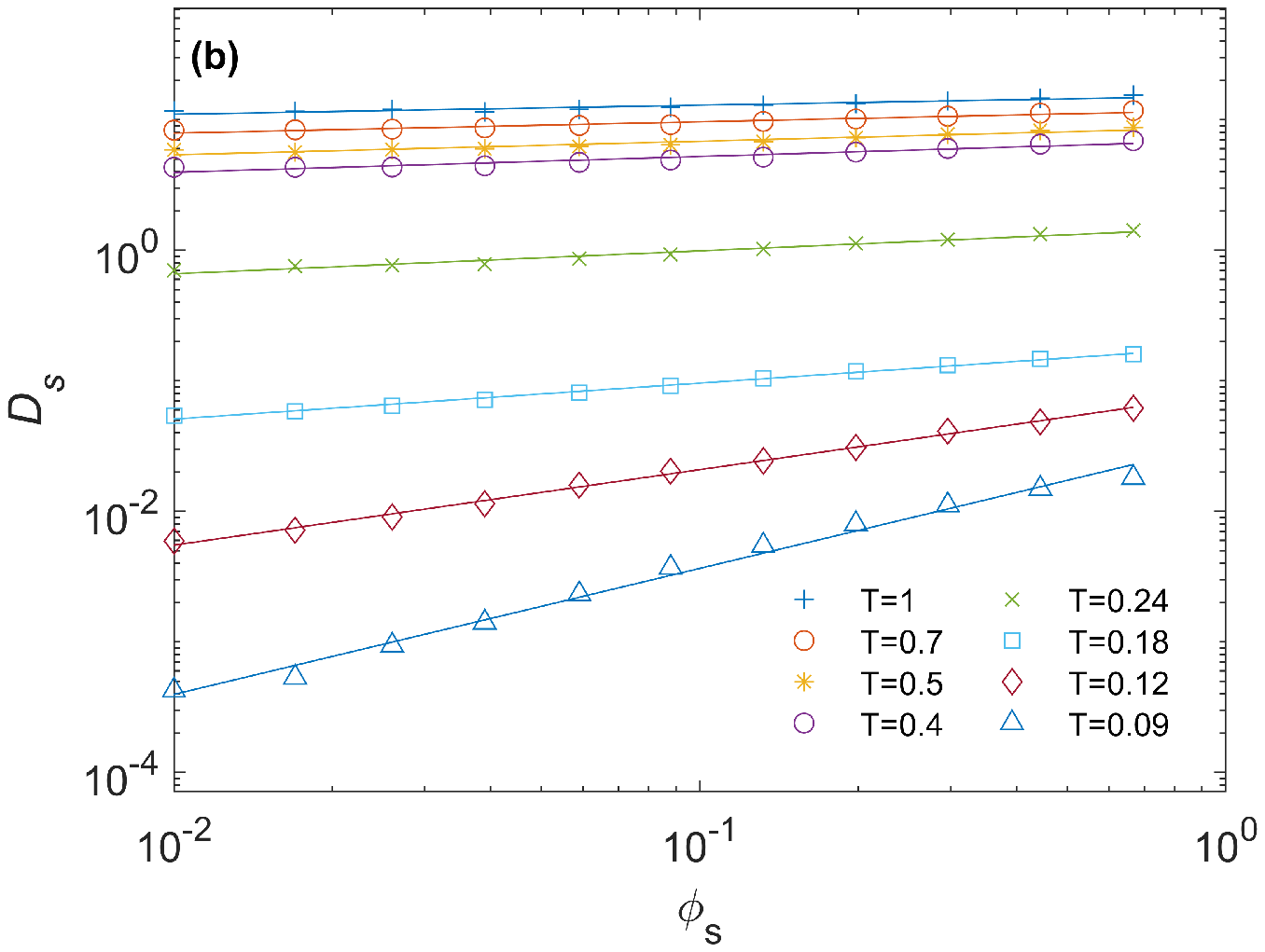}
		\caption{Diffusion coefficients $D_r$ of regular particles (a) and $D_s$ of swap-initiators (b) against the fraction $\phi_s$ of swap-initiators. }
		\label{fig:power_law}
	\end{figure}
    Figure \ref{fig:power_law}(a) shows $D_r$ as a function of $\phi_s$. The remarkably straight lines in the log-log plot show the power law
	\begin{equation}
      D_r \sim \phi_{s}^{\alpha}
      \label{Dr}
    \end{equation}
    for small $\phi_s $.
Similarly, Fig. \ref{fig:power_law}(b) plots the diffusion coefficient $D_s$ of the swap-initiators analogously defined. We observe a related power law: 
	\begin{equation}
		D_s \sim \phi_{s}^{\alpha-1}.
      \label{Ds}
    \end{equation}
It appears that these power laws cannot be inferred from existing theories of swap dynamics \cite{szamel2018,brito2018,ikeda2017}.

\begin{figure}[tb]
		\centering
		\includegraphics[width=\columnwidth]{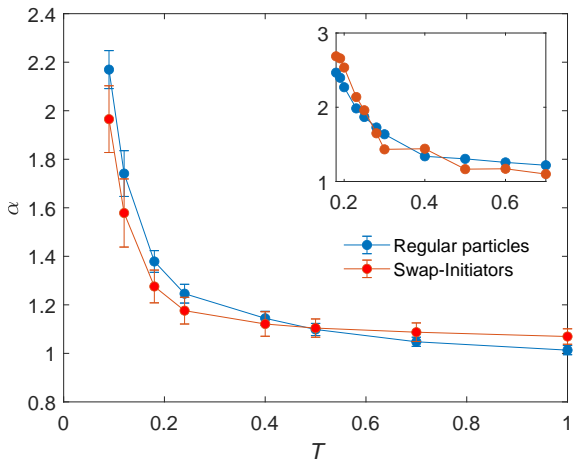}
		\caption{Scaling exponent $\alpha$ measured from regular particles (blue) and swap-initiators (red) against temperature $T$. Inset: Exponent $\alpha$ against T from Distinguishable Particle Lattice Model simulations. }
		\label{fig:alpha}
	\end{figure}

We first explain the relationship between the two power laws. 
For small $\phi_s$, swap-initiators are sparse so that they typically swap with regular particles. An equal number of swaps is thus shared between the entire population of the two species. The ratio of the swapping rates for regular particles against swap-initiators is hence inversely proportional to their population ratio $(1-\phi_s)/\phi_s  \simeq \phi_s^{-1}$. This implies 
$D_r/D_s \propto \phi_s$.  The exponents in \eqs{Dr}{Ds} thus differ by 1 and can be  denoted by $\alpha$ and $\alpha-1$ respectively.
Figure \ref{fig:alpha} plots $\alpha$ against $T$ from fitting $D_r$ and $D_s$ to \eqs{Dr}{Ds} respectively.
The reasonable consistency between values of $\alpha$ obtained from the two power laws 
supports our arguments.

The exponent $\alpha$ approaches 1 at high $T$ as observable from Fig. \ref{fig:alpha}.
In that case, $D_s$ is independent of $\phi_s$ and the swap-initiators are simply independent random walkers, indicating that thermal  motions readily overcome random particle interactions in the disordered system. More interestingly, $\alpha$ rises as $T$ decreases and exceeds 2 at $T \alt 0.09$.
Consider for example $\alpha = 2$ at  $T \simeq 0.09$. Based on elementary chemical kinetics, we suggest that pairs of nearby swap-initiators dominate the dynamics. These pairs have a density ${\sim}\phi_s^2$ so that the total swapping rate in the system is proportional to $\phi_s^2$. Distributing these swaps to individual initiators of population ${\sim}\phi_s$, we get $D_s \sim \phi_s$.
We expect that non-integral values of $\alpha$ are associated with crossover situations due to fluctuations in the dominant mobile group size. As $\alpha$ can exceed 2, our 
arguments may be applied to $\alpha$ equal to 3 or beyond corresponding to larger group of initiators which dominate the dynamics. 

\begin{figure}[tb]
		\centering
		\includegraphics[width=0.8\columnwidth]{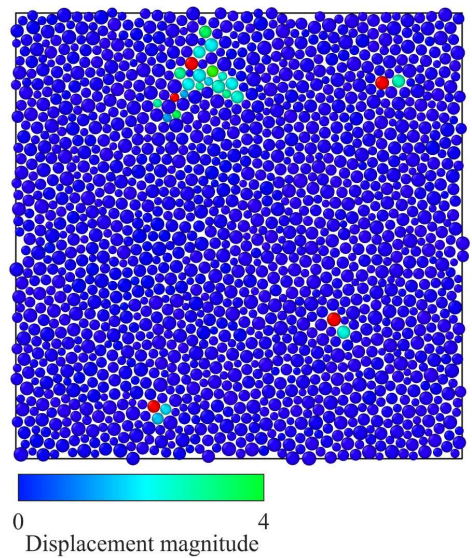}
		\caption{A 1600 particle system with $\phi_s$=0.003 at $T=0.08$ showing facilitation. Regular particles (colored according to their displacements during a time interval of 0.32) are more mobile when they are close to a pair of swap-initiators (red).}
		\label{fig:heterogeneity}
      \end{figure}

    The above picture of dynamics-dominating groups of defects, akin to the facilitation picture of glassy dynamics, was pioneered by the Fredrickson-Andersen model \cite{fredrickson1984} and further developed in numerous works \cite{garrahan2011review,keys2013,isobe2016,zhang2017}. The facilitation in our model is probabilistic, unlike the rigid rules in typical kinetically constrained models \cite{fredrickson1984,garrahan2011review}. Its relevance can be illustrated from real space displacement profiles.   \Fig{fig:heterogeneity} shows the displacement of a system at $T=0.08$ corresponding to $\alpha\simeq 2.1$. We have also taken a small $\phi_s$ so that individual groups of swap-initiators can be examined.
As a typical trend at such a low $T$, we can observe that regular particles close to a pair of swap-initiators are in general much more mobile than those next to isolated initiators. We corroborate this quantitatively in the SI by comparing systems in which swap-initiators are isolated, in pairs, or in triplets. 
      
\begin{figure}[tb]
		\centering

		\includegraphics[width=\columnwidth]{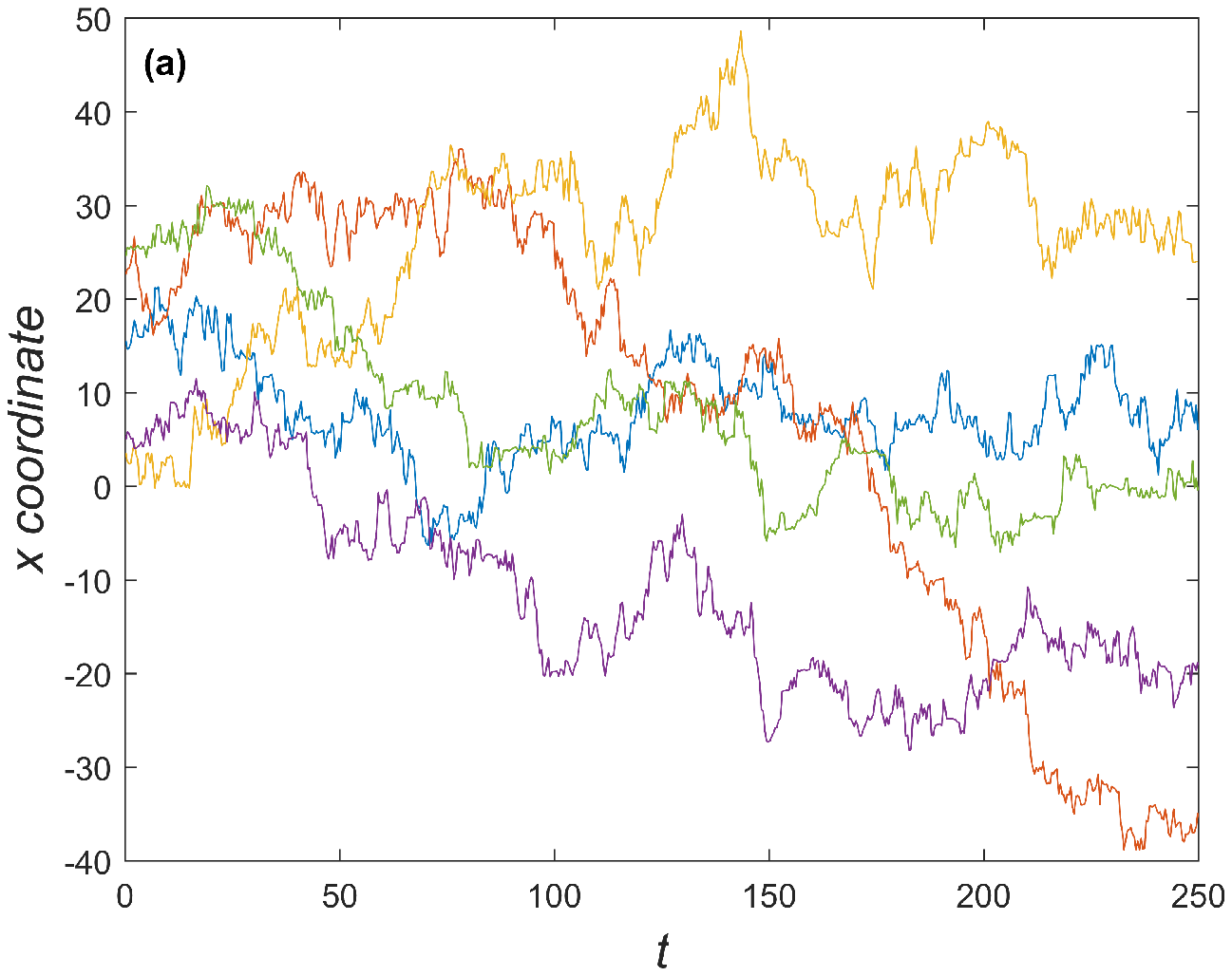}
		\includegraphics[width=\columnwidth]{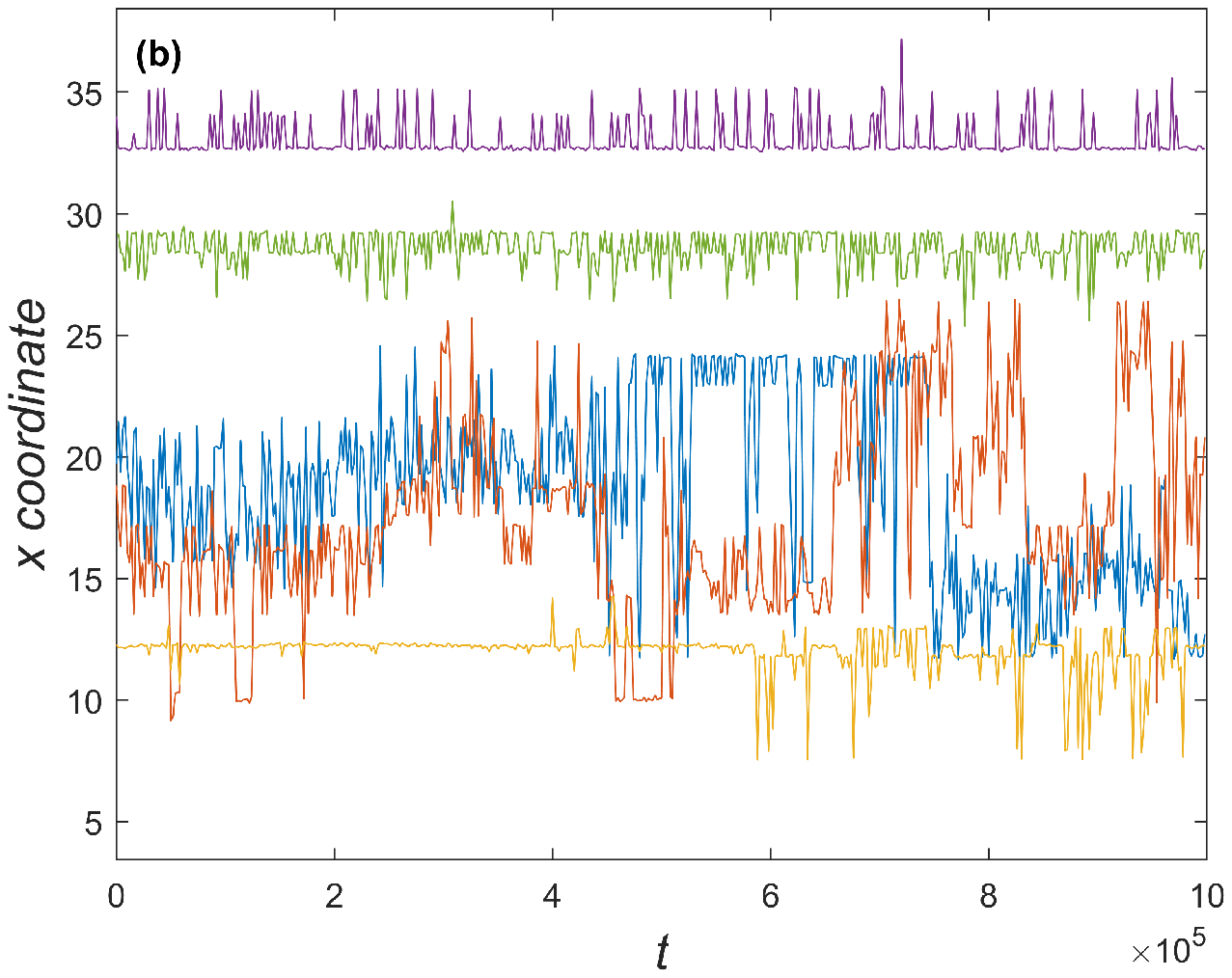}
		
    \caption{Plots of $x$-coordinates of all five swap initiators (randomly colored) in a system with 1600 particles for $\phi_s=0.003$ against time $t$ at (a) $T$=0.5  and (b) $T$=0.08. The coordinates are unwrapped with respect to the periodic boundary conditions for clarity. 
      In (a), motions are consistent with simple random walks. In (b), some swap initiators are strongly trapped, but two initiators close to each other are more mobile.
    }
		\label{fig:initiator}
	\end{figure}

      \Fig{fig:initiator} compares the position-time graphs of swap initiators at high and low $T$. It is clear that at high $T$, all swap initiators are mobile with motions consistent with random walks. In contrast, there are much stronger fluctuations at low $T$.  Some swap initiators are trapped for long duration within small regions. Importantly, they do not only vibrate as one may naively expect for caged particles. Instead, they swap frequently, albeit only back-and-forth leading to little net movements. They are thus essentially caged, but in a more general sense with back-and-forth swaps and vibrations. We also observe two other swap initiators which are much more mobile. We have checked in this and other examples that the mobile initiators are mainly those in groups of two or more. Some of these coupled groups are completely mobile and move over long distances. When they reach trapped swap initiators, partners may be exchanged and thus no initiator is permanently trapped in a sufficiently large system.
      
	Another intriguing feature of our model is that the regular particles exhibit remarkably glass-like dynamics at small $\phi_s$. In this regime,
dynamics is predominantly induced by only a small population of frequently swapping initiators. 
Yet, the regular particles demonstrate typical glassy behaviors. These include a mean squared displacement exhibiting a plateau, a two-step decay of the self-intermediate scattering function with a stretching exponent decreasing with $T$, a Stokes-Einstein violation, and a peak in a time-dependent four-point susceptibility with a height increasing as $T$ decreases (see SI).


More directly, real-space features typical of glass formers can also be observed and intuitively understood. Specifically, Fig. \ref{fig:heterogeneity}  shows dynamic heterogeneity
revealed as a cluster of regular particles with a much higher mobility than the others. This high mobility results simply from the proximity to a facilitated pair of swap initiators.
  Another important real-space feature is string-like motions, involving  strings of particles displacing their preceding neighbors \cite{glotzer1998}. These are revealed as strings of mobile regular particles in \figs{fig:system}{fig:heterogeneity}. Some mobile regular particles in \fig{fig:heterogeneity} seem to form compact geometries, which indeed can be broken down into strings at shorter time intervals.  A more informative illustration is provided by particle trajectories, where   individual trajectories of groups of particles nicely connect to form strings  (see SI). We observe striking resemblance of these strings to those in, for example, experimental glassy colloidal systems \cite{yip2020}.
Unlike in realistic glass, string-like motions in our model can be trivially understood. Each is simply caused by a few consecutive local swaps of a swap initiator, leaving behind a linear trail of displaced particles. Their trajectories thus align to form a string.

Many important properties of glass are captured by lattice models \cite{garrahan2011review}. We have recently proposed a distinguishable particle lattice model (DPLM) \cite{zhang2017}, which exhibits a wide range of glassy phenomena (see e.g. \cite{lulli2020,lee2020}). Generalizing the DPLM to incorporate swap, we have reproduced both power laws with exponents showing similar $T$ dependence (see inset in \fig{fig:alpha}
and SI).
  

\section{Discussion}
We have shown that introducing a density $\phi_s$ of swap initiators and implementing local particle swaps, simple power-law relations between diffusion coefficients, the most fundamental dynamic measures, and $\phi_s$ are established. The scaling exponents depend non-trivially on temperature. These are highly specific hallmark features fundamental to swap dynamics in glass formers. Power laws play key roles in theoretical descriptions of many physical systems and techniques to tackle them are abundant \cite{plischke1994}. Incorporating them into existing \cite{szamel2018,brito2018,ikeda2017} and future theories of swap dynamics should be important in scrutinizing and perfecting them. In addition, we have shown that the power laws can be reproduced using the DPLM. A theoretical description of swap with the power laws appears readily achievable, as lattice models are in general much more tractable analytically than MD systems  \cite{garrahan2011review,lam2018tree}.

We have found that glassy dynamics is exhibited by regular particles at small $\phi_s$. This is a highly nontrivial finding because in contrast to realistic glass formers in which all particles in-principle can move spontaneously, particle motions here are mainly induced by a sparse population of swap initiators.
In our opinion, the regular particles constitute the simplest molecular model of glass, as motions are clearly known to be caused by and localized around swap initiators.
It is so simple that dynamic heterogeneity and string-like motions are trivially understandable as explained above. A theory for swap dynamics at small $\phi_s$ should be highly inspiring, if not directly applicable, for a quantitative description of glassy dynamics.


We have argued that the power laws in the non-trivial regime with  $\alpha>1$ result from elementary chemical kinetics and relate to facilitation \cite{garrahan2011review}. 
It can further be explained intuitively as follows. Consider, e.g. $\alpha \simeq 2$. Although, a swap-initiator can in-principle swap with all its nearest neighbor associated with various energy costs, only some of them can be energetically favorable at such a low $T$. If these neighbors do not percolate throughout the whole system, the initiator will be trapped to move back-and-forth only within a few sites defined by the rugged energy landscape. Isolated initiators thus tend to have a low mobility at a sufficiently low $T$. Importantly, besides being affected by the energy landscape, motions of swap initiators indeed also perturb the  landscape as particle arrangements along their pathways are altered. Therefore, if a swap initiator happens to move close to another one,  the landscape experienced by the latter will be perturbed and this may unlock previously unfavorable swaps. More generally, both initiators perturb the energy landscape of each other and provide additional swapping possibilities. This mutual facilitation can enable both to move far away in a dynamically coupled way. Analogous facilitation based on void-induced dynamics has been explained in detail previously \cite{lam2017,lam2018tree, deng2019}.

The swap-initiators at small $\phi_s$ are mobile point defects in the system of regular particles. Our results thus show that many features of glassy dynamics can well be realized by defect-induced motions. An important question is: are there analogous dynamics-dominating defects in realistic glasses? In close association with the free-volume theory, a major candidate is void or, more generally, a fragmented version called quasi-void, which has been recently identified in colloid experiments via a reversible transformation into a vacancy at a glass-crystal interface \cite{yip2020}. In this picture of void-induced dynamics, a particle hopping into a nearest-neighboring void, leaving another void behind, can be equivalently described as a local swap between a particle and a void. This is fully analogous to a local swap between a regular particle and a swap-initiator. 
The two formalisms are thus intimately related. 
We have recently proposed a description of such void-induced glassy dynamics  \cite{lam2018tree, deng2019}, which will be applied in the future in attempt to account for the present results quantitatively.


\section{Conclusion}
We have introduced a partially-swapping system and found simple power laws relating diffusion coefficients to the density $\phi_s$ of swap-initiating particles. The exponents of the power laws depend nontrivially on temperature. These observations have not been predicted by existing theories of swap, but can be explained by facilitation and are reproduced with a lattice model. In addition, the system exhibits remarkably typical glassy dynamics at small $\phi_s$, implying that main characteristics of glass formers can be defect induced.

We are thankful for the many helpful discussions with H.B Yu and M. Lulli. This work was supported by General Research Fund of Hong Kong (Grant 15303220), National Natural Science Foundation of China (Grant 11974297 and 12174079), GuangDong Basic and Applied Basic Research Foundation (Grant  2214050004792) and Shenzhen Municipal Science and Technology projects (Grant 202001093000117).


	\bibliography{ms}

\end{document}


\title{Supplementary Information}
	\author{Gautham Gopinath$^1$}
	\author{Chun-Shing Lee$^1$}
	\author{Xin-Yuan Gao$^1$}
	\author{Xiao-Dong An$^1$}
	\author{Chor-Hoi Chan$^2$}
	\author{Cho-Tung Yip$^2$}
	\author{Hai-Yao Deng$^3$}
	\author{Chi-Hang Lam$^1$}
	\address{
		$^1$Department of Applied Physics, Hong Kong Polytechnic University, Hung Hom, Hong Kong, China \\
		$^2$Department of Physics, Harbin Institute of Technology, Shenzhen 518055, China\\
		$^3$School of Physics and Astronomy, Cardiff University, 5 The Parade, Cardiff CF24 3AA, Wales, UK
	}
	\maketitle

	\section{Details on Molecular dynamics simulations with partial swap}

\subsection{Model}
    In our main simulations, we study a two-dimensional system with $1024$ particles in a square region of size $32 \times 32$, leading to a unit particle number density.      
We consider continuously polydisperse repulsive particles.
The diameter $\sigma_i$ of particle $i$ is sampled from a uniform distribution in the range [0.682, 1.318], which gives a mean diameter $\bar{\sigma}=1$. The standard deviation of the diameter is $0.18$, characterizing the degree of polydispersity of the system. Noting that successful swap attempts usually occur between particles of similar radii, the uniform diameter distribution allows comparable successful swapping rates of both small and large particles and similar distributions have been studied previously \cite{fernandez2007}.

Two particles with diameters $\sigma_i$ and $\sigma_j$ separated by a distance $r$ interact through a soft repulsive potential given by:
	\begin{equation}
		V_{ij} =v_o\left(\frac{\sigma_{ij}}{r}\right)^{12} 
      \end{equation}
      where $v_0=1$ and
	\begin{equation}
      \sigma_{ij}=\frac{(\sigma_i+\sigma_j)}{2}(1-\epsilon|\sigma_i-\sigma_j|).
    \end{equation}
An interaction cutoff distance of $1.25\sigma_{ij}$ is applied beyond which the interaction is truncated. We use a non-additivity parameter $\epsilon = 0.2$  that promotes particles with a larger size difference to stick close to one another, suppressing fractionation within the system. This combination of size polydispersity and non-additivity has been shown to produce highly stable glass formers \cite{berthier2019,ninarello2017}.  The mean diameter $\bar{\sigma}$ defines the unit of length in our simulations, and $v_0$ defines the unit of energy. The unit of time then becomes $\bar{\sigma}\sqrt{m/v_0}$, where $m=1$ is the mass of each particle. We take the Boltzmann constant $k_B=1$ so that the unit of temperature is also $v_0$. All these quantities are set to 1, leading to dimensionless units adopted in this work.
	
\begin{figure}[tb]
 	\includegraphics[width=\columnwidth]{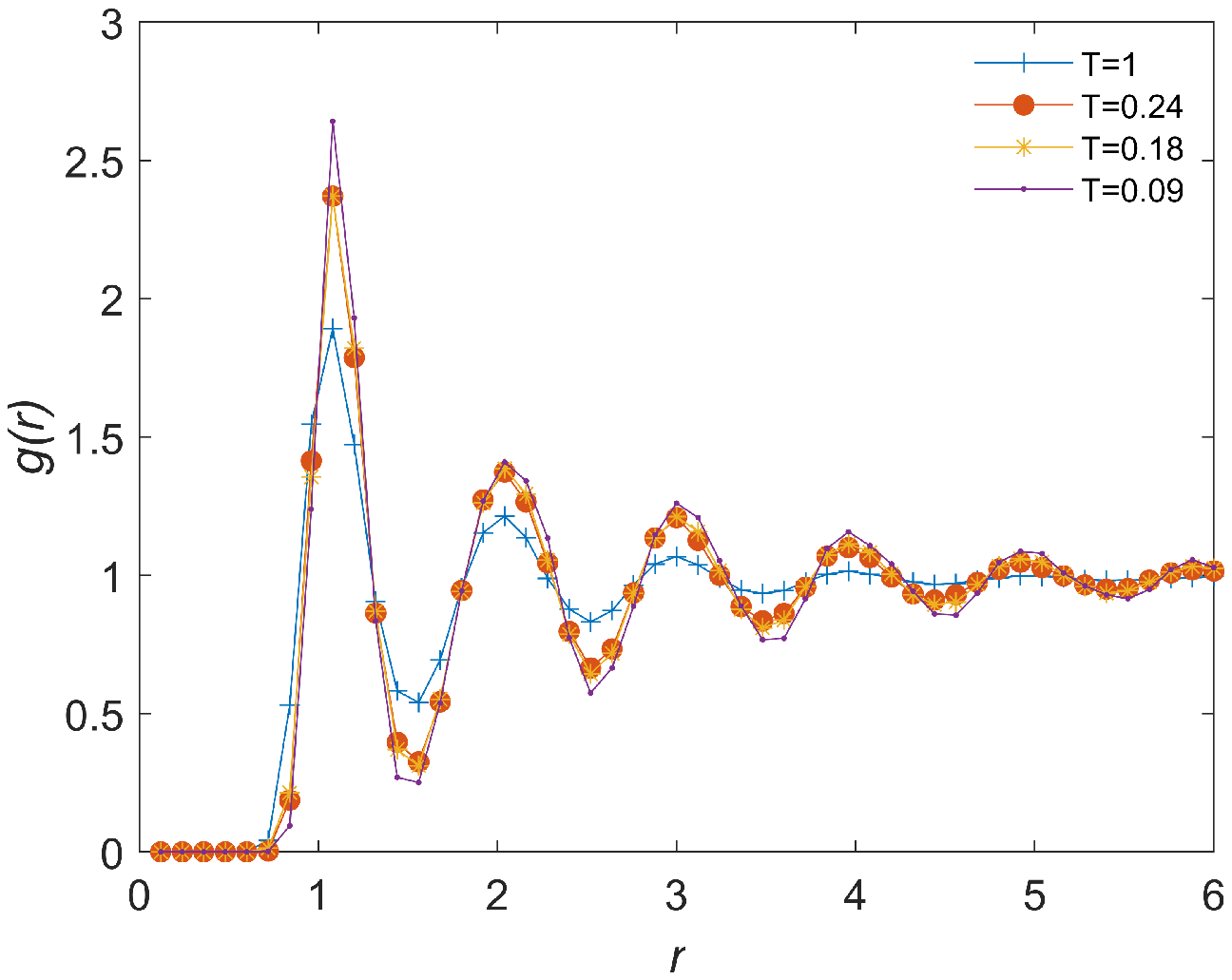}
 	\caption{Particle pair distribution functions $g(r)$ at various temperatures.}
 	\label{fig:rdf}
 \end{figure}

      Our molecular dynamics (MD) simulation is a hybrid one with periodic swap Monte Carlo steps, analogous to related hybrid approaches.
The time-step for the Verlet integration of the MD process is $\Delta t=0.001$. 
In our low temperature simulations for example, after each MD time window of width 0.2, we perform 200,000 swap attempts, leading to a swap attempt rate of $\mu=10^6$ (other $\mu$ has been used, as mentioned below).
 We adopt local swaps which exchange the positions of two particles within a swap cutoff distance $R_{swap}=1.5$, which is close to the first minimum in the particle pair distribution function. We calculate the particle pair distribution functions $g(r)$ as shown in Fig. \ref{fig:rdf} defined as follows:
 	
 	\begin{equation}
 		g(r) = \frac{\langle \rho(\boldsymbol{r_0}) \rho(\boldsymbol{r_0}+\boldsymbol{r}) \rangle}{{\langle \rho(\boldsymbol{r_0}) \rangle }^2}
 		\label{gr}
 	\end{equation}
 	where $\rho$ is the particle density, $|\boldsymbol{r}|=r$ and the average is over positions $\boldsymbol{r_0}$.

A key feature in our approach is a partial swap algorithm.
  Before a simulation starts, we randomly select a fraction $\phi_s$ of particles in the system, referred to as {\it swap-initiators}. Their number density is thus also $\phi_s$. Only these particles can perform swaps with themselves or with other regular particles. Regular particles cannot swap directly among themselves. 

\subsection{Simulation methods}
  Each swap attempt at temperature $T$ employs the below algorithm, which follows detailed balance: 
	\begin{enumerate}
		\item Randomly select two particles with uniform probability.
		\item If neither of the selected particles is a swap-initiator OR the particle separation is greater than $R_{swap}$, reject the attempt.
		\item Otherwise, accept the swap with probability $\min \{ 1, e^{-\frac{\Delta E}{k_BT}} \}$, where $\Delta E$ is the energy difference between the system after and before the swap. 
	\end{enumerate}
    According to this algorithm, a good majority of swaps are rejected at step 2
and do not involve an intensive energy calculation. In particular, if one takes $\phi_s = 1$ and $R_{swap}=\infty$, this algorithm reduces to the more common non-local algorithm with full swapping \cite{ninarello2017}.

All simulations are performed using LAMMPS \cite{thompson2021lammps}. As LAMMPS does not handle continuous polydispersity natively, it is approximated by a discrete set of 128 diameter values and python scripts have been written to define all these particle types and generate a large table of all possible interactions to be input to LAMMPS. Swaps are performed using the Monte Carlo package of LAMMPS \cite{sadigh2012}. However, it only provides non-local swap with full swapping. We have performed source code modifications to allow for local and partial swap as well as to enable a single simplified command for swapping among all particle types of various radii. In our main simulations used for quantitative measurements of, for example, diffusion coefficients, results at each $T$ and $\phi_s$ are typically averaged over 30 independent simulations which run for time $t\simeq10^5$. Pictures of the system in the main text are produced using OVITO \cite{ovito}.

\subsection{MSD and swap-dominating regime}
\begin{figure} [tb]
	\includegraphics[width=\columnwidth]{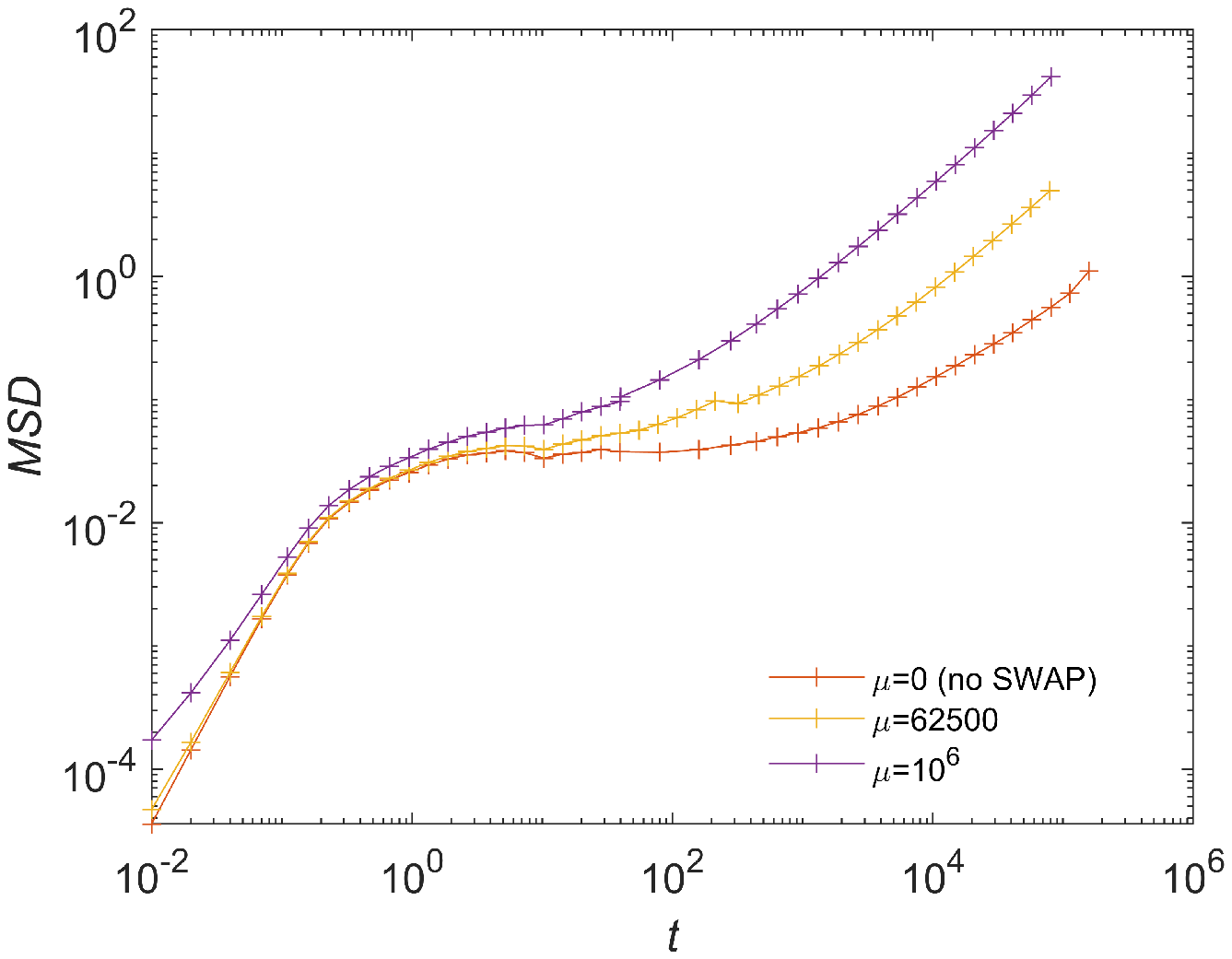}
	\caption{Mean-squared displacement of regular particles for various swapping rates $\mu$. We take $T=0.18$ and $\phi_s=0.01$.}
	\label{fig:msdrate}
\end{figure}

Many of our measurements are based on particle mean-squared displacement defined by 
MSD $= \langle |\textbf{r}_i(t)-\textbf{r}_i(0)|^2 \rangle$, where $\textbf{r}_i(t)$ is the position of particle $i$ at time $t$. The MSD for regular particles
is shown in \fig{fig:msdrate} for selected swapping rates $\mu$. We have included the case of  $\mu=0$ corresponding to the absence of swap, which indicates the significance of the MD dynamics at the given $T$. The diffusion coefficient $D_r$ of the regular particles is then calculated from $D_r = \text{MSD} /{4t}$ measured at long time $t$.

All the main simulations reported in this work adopt sufficiently large values of $\mu$ so that the dynamics is swap-dominated. For $T=0.09$ to 0.2 we adopt $\mu=10^6$. For $T>0.2$ we use $\mu=2.5\times10^7$, such that the diffusion coefficient $D_r$ of the regular particles with the partial-swap dynamics is at least 10 times higher than that without swap.
This also means that the MSD at long time must be enhanced by swaps by at least 10 times.
For example, from \fig{fig:msdrate}, the rate $\mu=10^6$ brings the system well within the swap-dominating regime for that given $T$ and $\phi_s$, while $\mu=62,500$ does not. 

\begin{figure} [tb]
	\includegraphics[width=\columnwidth]{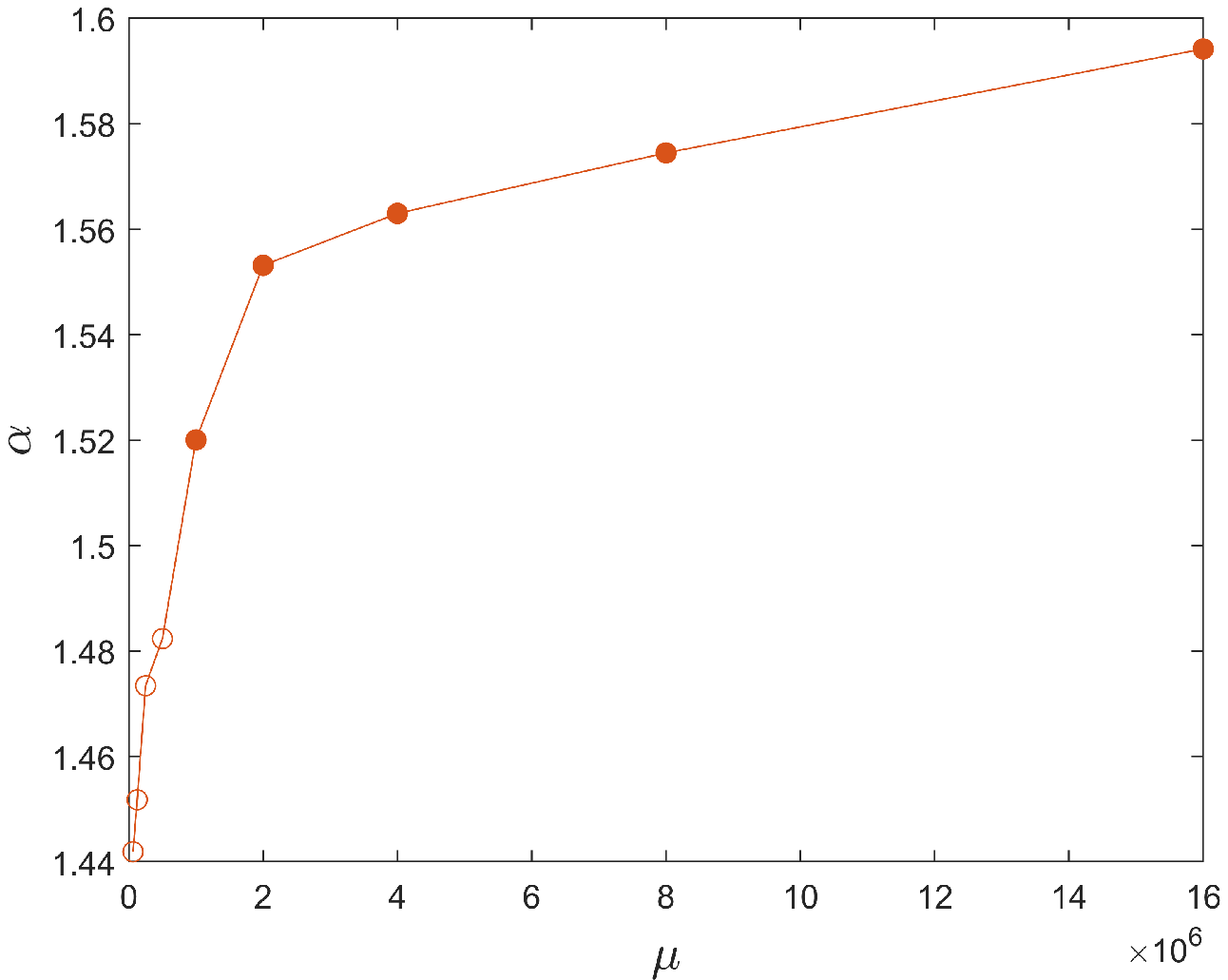}
	\caption{Power-law exponent $\alpha$ for regular particles against the swapping rate $\mu$. It is small at small $\mu$ (open circles) and becomes relatively independent of $\mu$ at large $\mu$ in the swap-dominating regime (solid circles). We take $T=0.18$. }
	\label{fig:alpha_rate}
\end{figure}

Furthermore, we show here that once $\mu$ is large enough, the variation of the power-law exponent $\alpha$ for various $\mu$ is small.    Figure \ref{fig:alpha_rate} plots this relationship.  At small $\mu$, the exponent $\alpha$ measured is relatively small. This is because MD dynamics contributes significantly to the MSD so that varying $\phi_s$ has a reduced impact.
For larger $\mu\agt 2 \times 10^6$  satisfying our  swap-dominating criterion, it then shows negligible dependence on $\mu$.
In fact, $\alpha$ varies very little; by less than 5 percent as $\mu$ changes by one order of magnitude. This indicates that, in the swap-dominating regime, $\mu$ is not a primary determinant of $\alpha$, in contrast to $T$ for instance. 

\subsection{Measurements over all particles}
\begin{figure}[tb]
		\centering
		\includegraphics[width=\columnwidth]{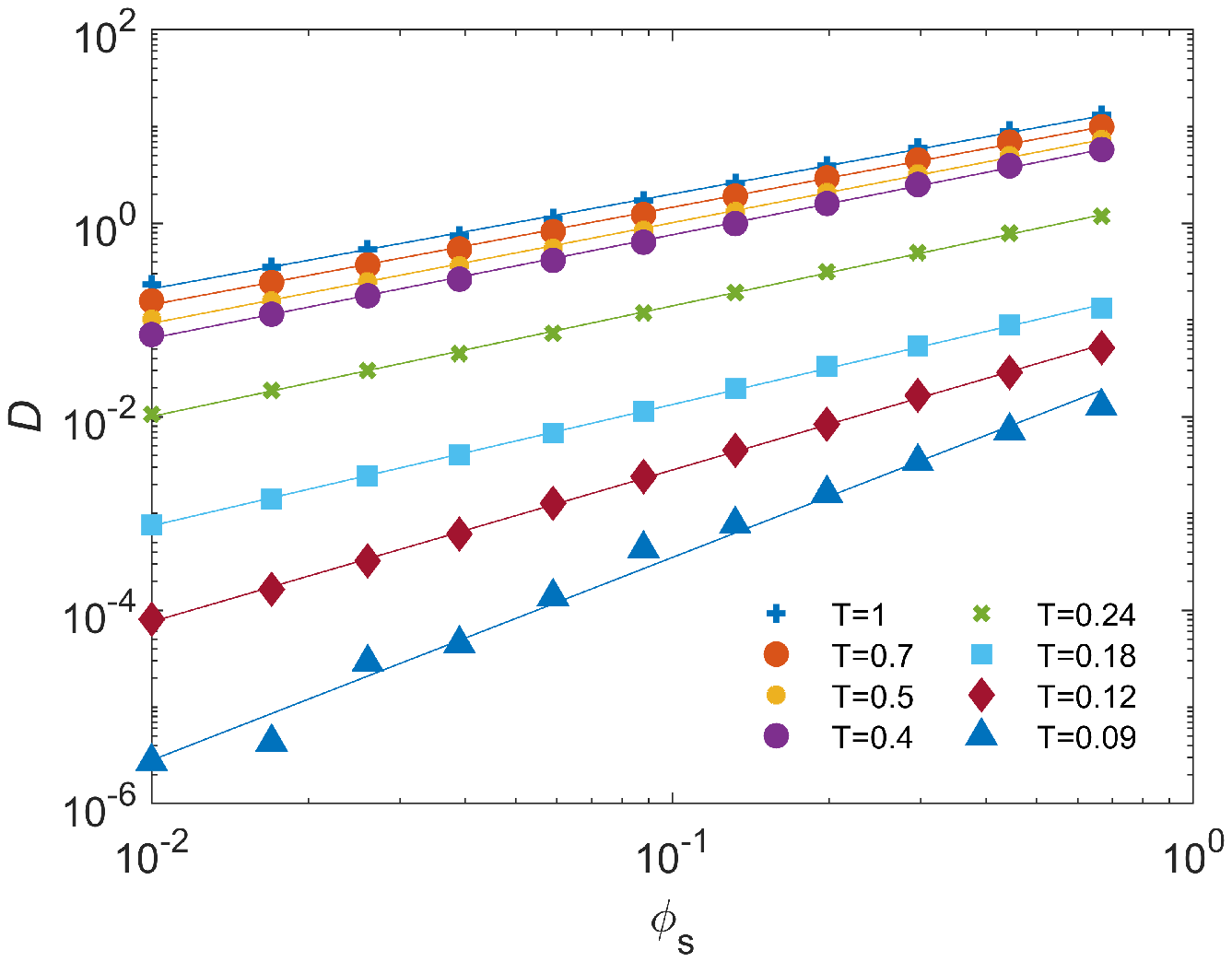}
		\caption{Diffusion coefficient $D$ against swap-initiator fraction $\phi_s$.} 
		\label{fig:power_law_all}
	\end{figure}

\begin{figure}[tb]
		\centering
		\includegraphics[width=\linewidth]{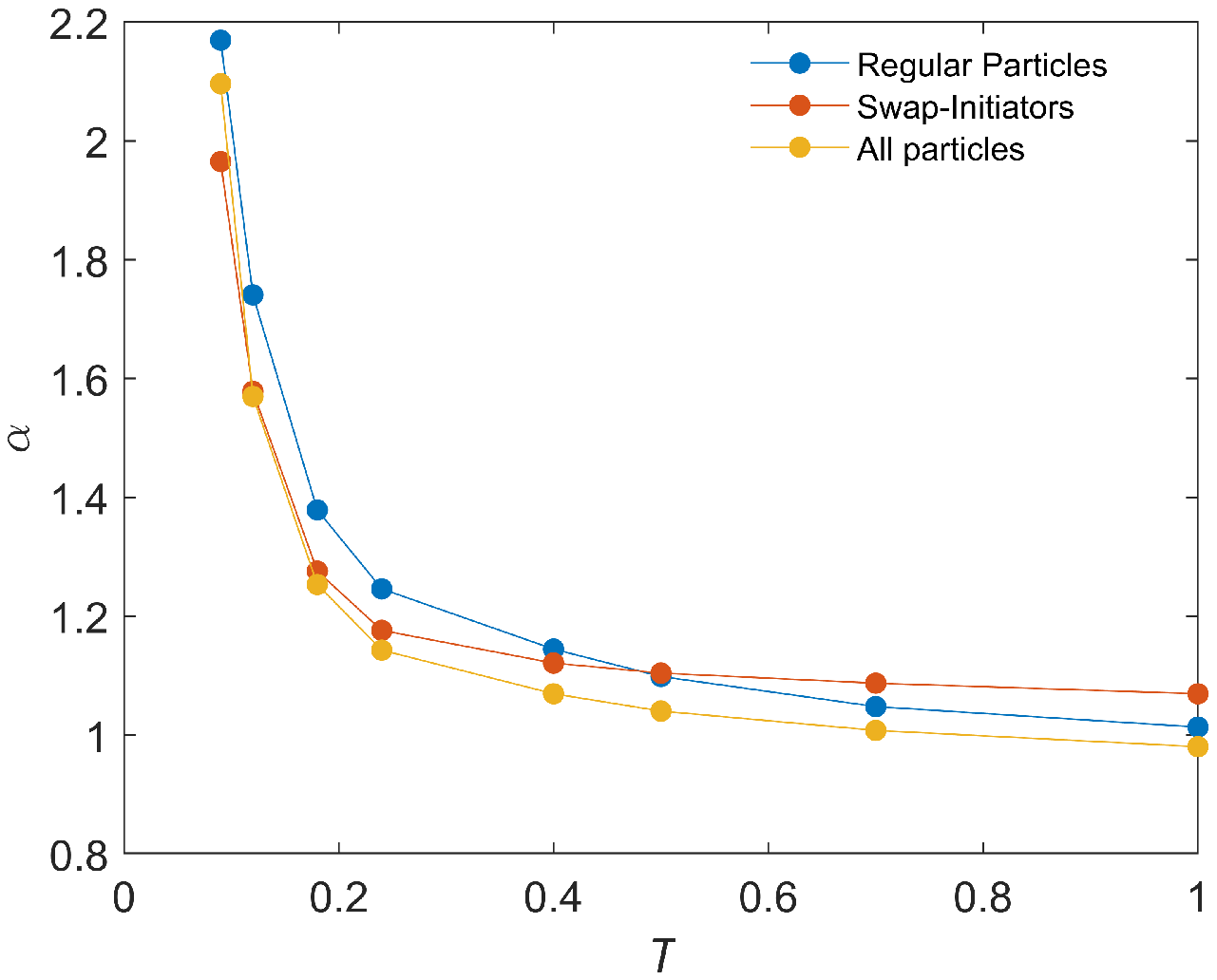}
		\caption{Power-law exponent $\alpha$ against temperature $T$ for all particles, regular particles and swap-initiators.}
		\label{fig:alpha_all}
	\end{figure}

We have reported in the main text the diffusion coefficients for regular particles and swap-initiators individually. For completeness, Fig. \ref{fig:power_law_all} plots the diffusion coefficient $D$ against $\phi_s$ measured for \textit{all} particles in the system.  It shows a power law resembling that for regular particles:
	\begin{equation}
      D \sim \phi_{s}^{\alpha}.
      \label{D}
    \end{equation}
This similarity is because of the abundance of the regular particles at small $\phi_s$ compared to the swap-initiators so that they dominate the overall dynamics.     
Figure \ref{fig:alpha_all} plots values of $\alpha$ measured from all particles, regular particles, and swap-initiators, which are all consistent with each other. 
	
    \section{Facilitation among swap-initiators}

\begin{figure}[tb]
    		\includegraphics[width=\columnwidth]{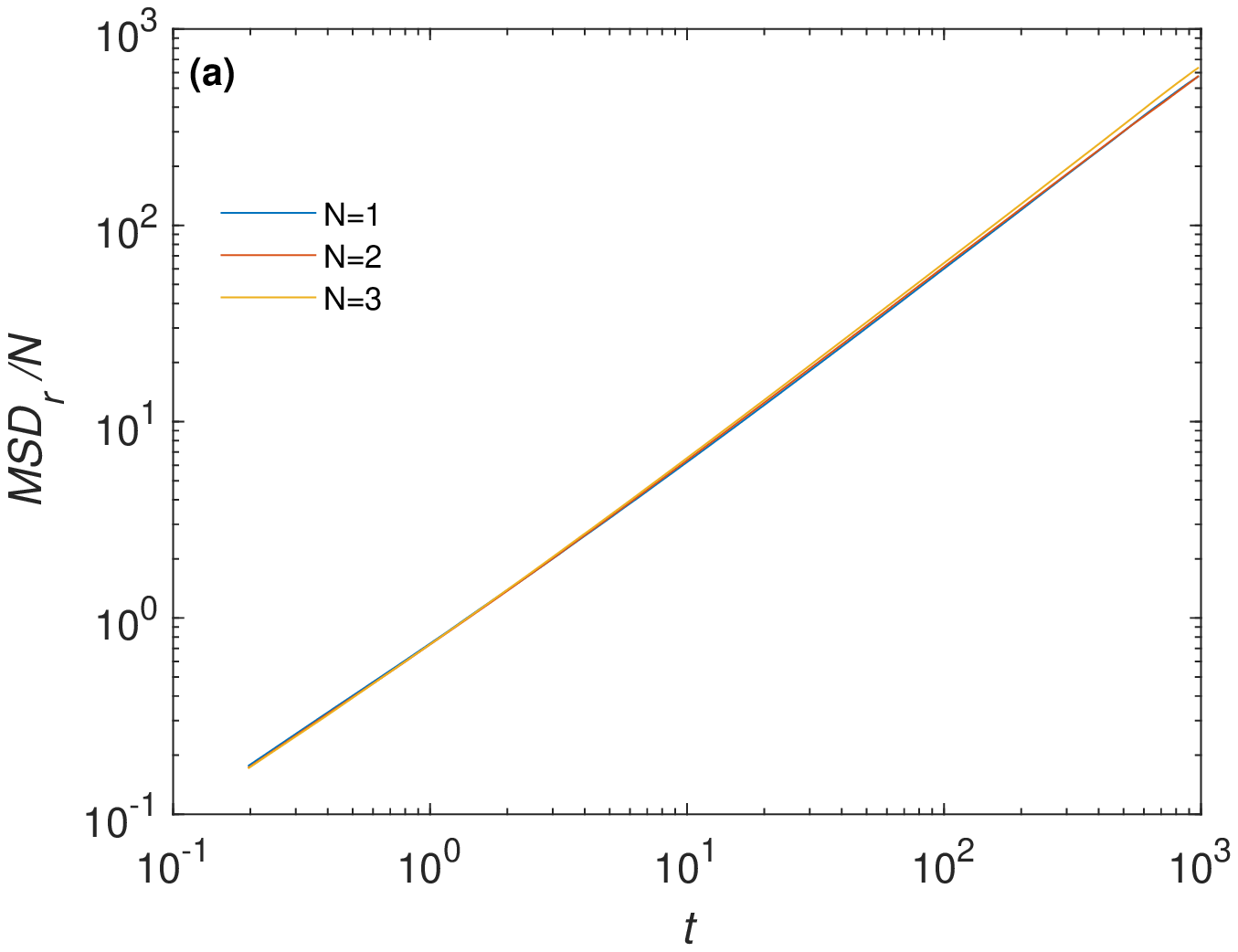}
    		\includegraphics[width=\columnwidth]{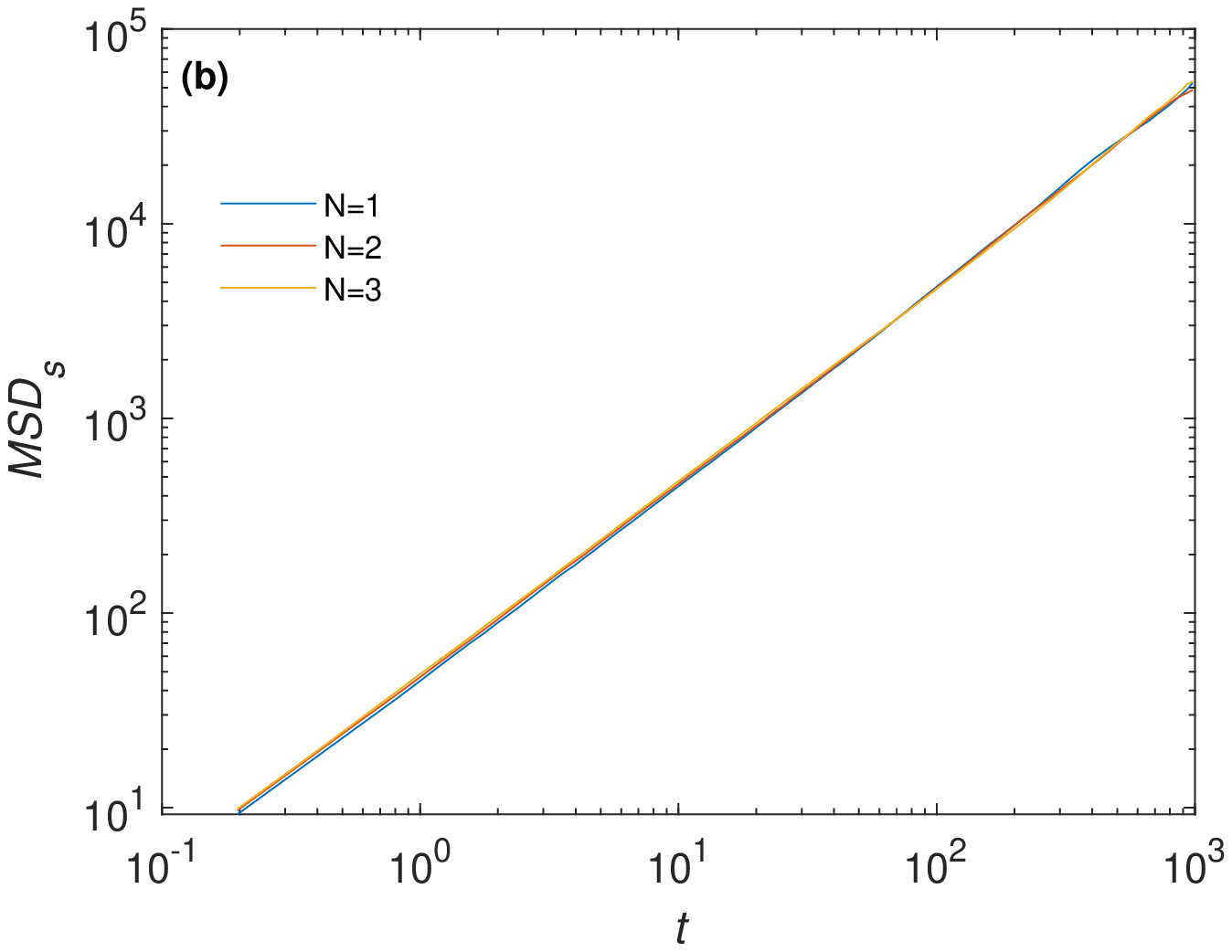}
    		\caption{(a)Normalized mean-squared displacements $\text{MSD}_r/N$ of regular particles and (b) mean-squared displacements $\text{MSD}_s$ of swap-inititiators in a small 10x10 system with N swap initiators at T = 0.5. At this temperature $\alpha \simeq 1.1$. }
    		\label{fig:mobility_highT}
    	\end{figure}

\begin{figure}[tb]
    		\includegraphics[width=\columnwidth]{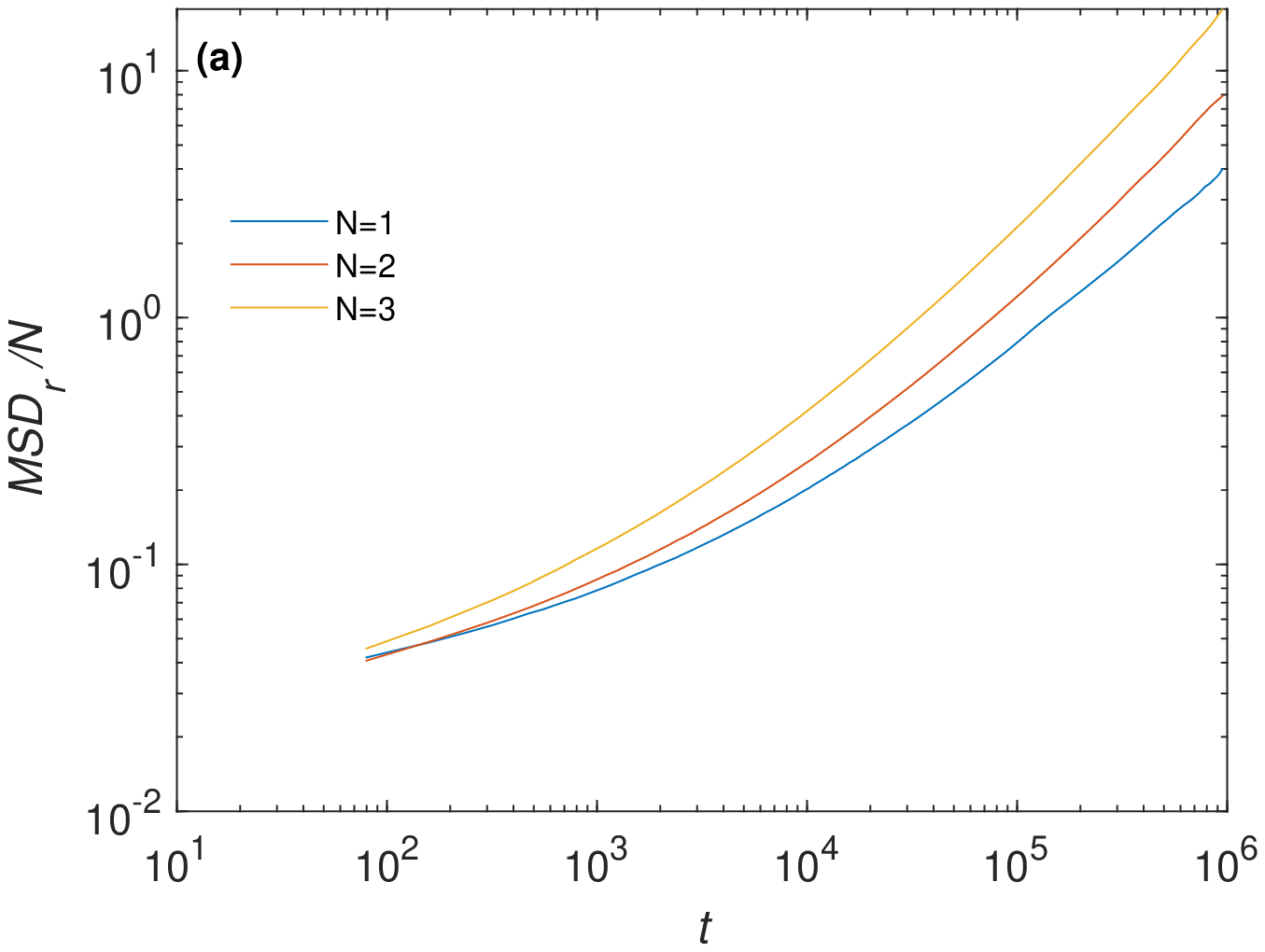}
    		\includegraphics[width=\columnwidth]{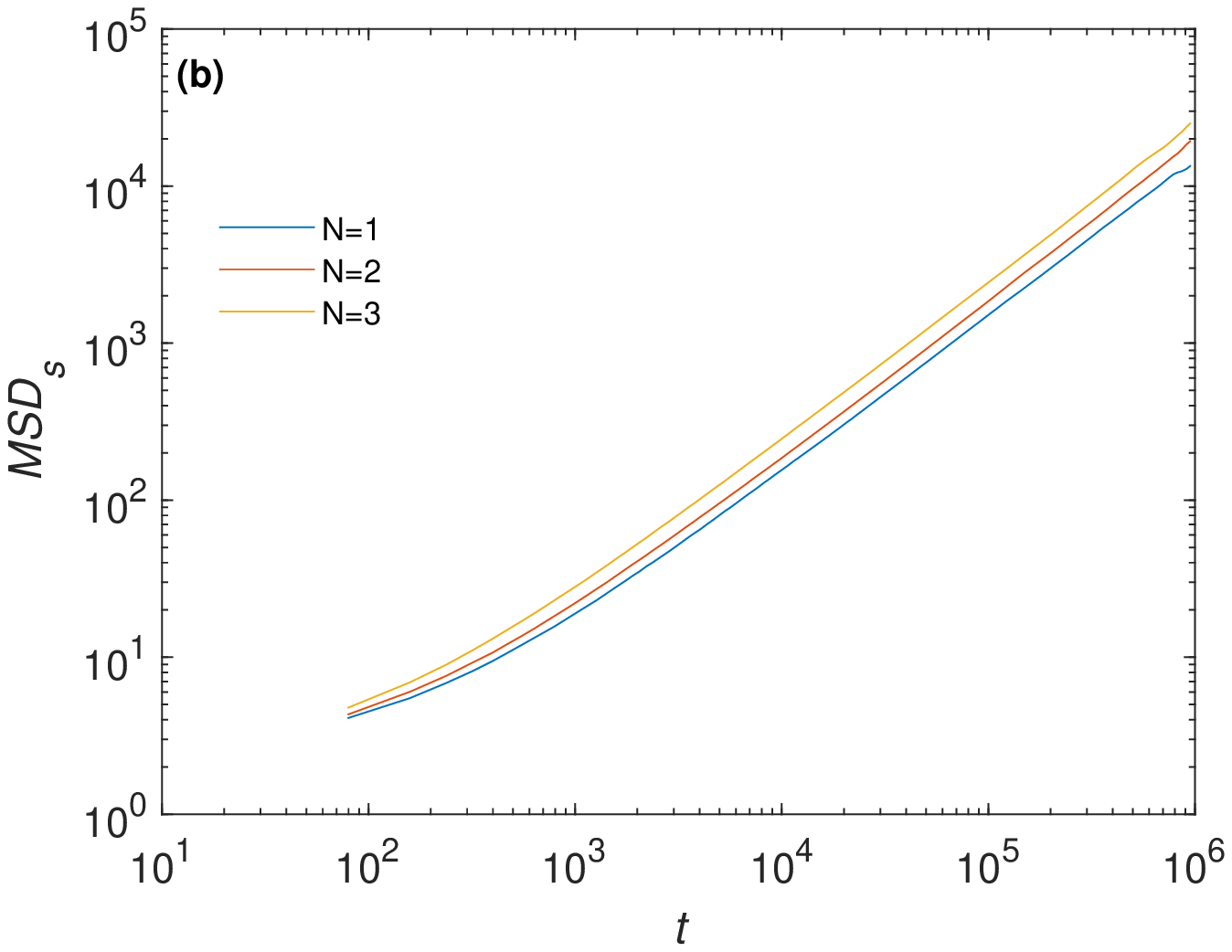}
    		\caption{Similar to \fig{fig:mobility_highT} but for T = 0.08 where $\alpha \simeq 2.1$.}
    		\label{fig:mobility_lowT}
    	\end{figure}

    	To further quantify the facilitation present in the system, we simulate smaller 10x10 systems as these allow swap-initiators placed together to remain close to one another during the simulation. (Larger systems give qualitatively similar but quantitatively weaker effects as swap-initiators are often observed to separate from each other and cease facilitating each other's motions). Thus, we measure the MSD of regular particles (MSD$_r$) and swap-initiators (MSD$_s$) for the cases where there are N=1, N=2, and N=3 swap-initiators placed randomly in the system at the beginning of a simulation. We average the results over 50 independent simulations each lasting for time $t=10^3$ and $7 \times 10^5$ for $T=0.5$ and $T=0.09$ respectively. The results are shown for a high temperature (where we would expect no facilitation) and a low temperature (where we would expect strong facilitation) in Figs \ref{fig:mobility_highT} and \ref{fig:mobility_lowT}.
    	
    	The MSD curves of the regular particles have been normalized by N to account for the trivially more swapping events induced directly by more swap-initiators even in the absence of facilitation. One can see that both MSD$_r/N$ and MSD$_s$ are nearly identical for a high temperature for which $\alpha \simeq 1.1$ according to Fig. 3 in the main text. In contrast, those at a low temperature where $\alpha \simeq 2.1$ are distinct. The larger values for the N = 2 and 3 cases indicate the presence of facilitation.

	\section{Glassy dynamics induced by a low density of swap-initiators}
	
	\begin{figure}[tb]
		\centering
			\includegraphics[width=\columnwidth]{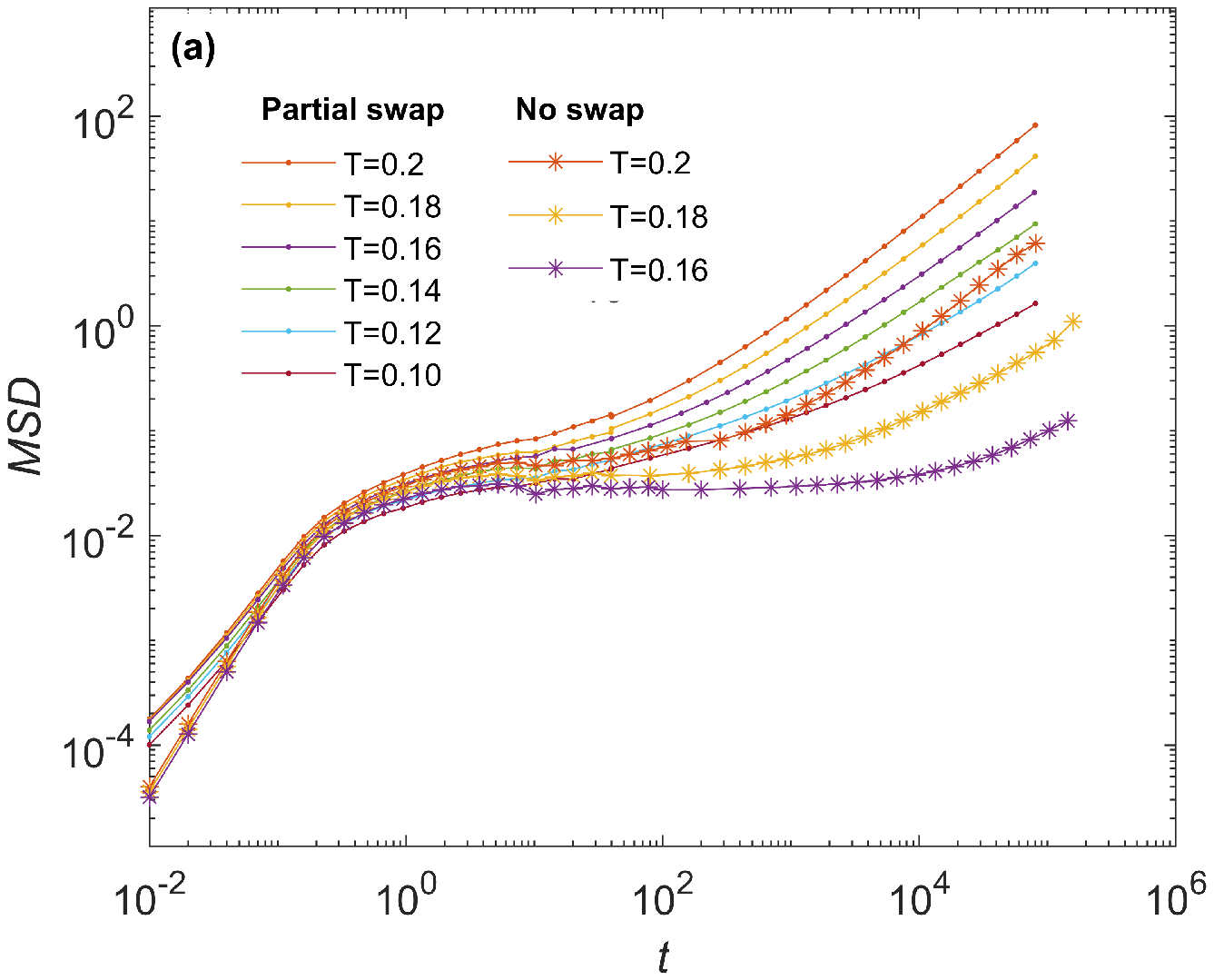}
			\includegraphics[width=\columnwidth]{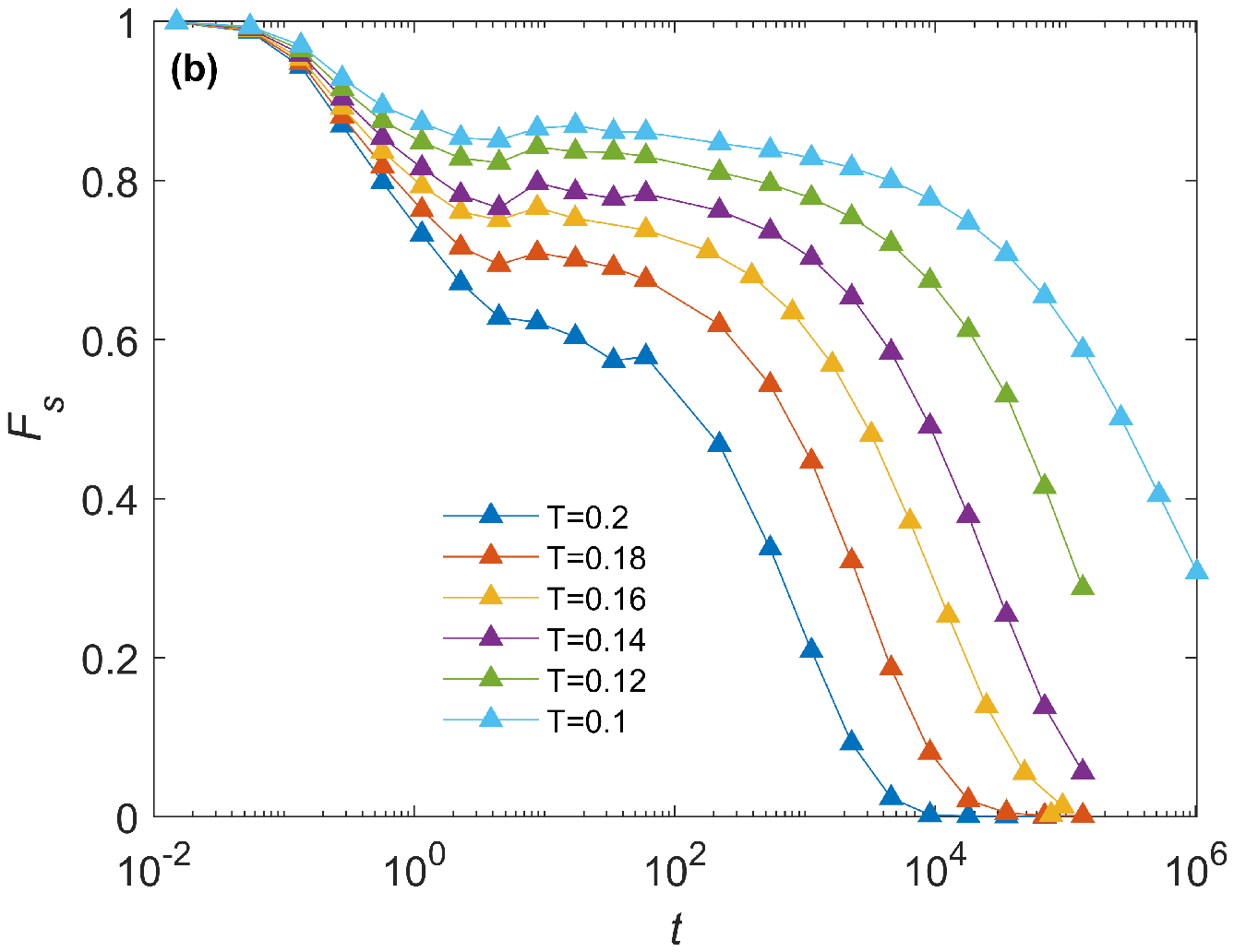}

		\caption{(a) Mean-squared displacement and (b) self-intermediate scattering function against time at various temperatures $T$ for $\phi_s$=0.01. Results in (a) are compared to those without swap.   }
		\label{fig:msd_sisf}
		
	\end{figure}
    
	Remarkably, the subsystem formed by the regular particles displays typical glassy behaviors at a small fraction $\phi_s$ of swap-initiators. With few swap-initiators, the regular particles nearly fill the space and model a typical material system by themselves.  As local swaps avoid long jumps, their dynamics resembles that in realistic particle systems. We focus on $\phi_s$ in the order of 0.01 but other small values are expected to generate qualitatively similar results. 


\Fig{fig:msd_sisf}(a) shows the
MSD at time $t$ for regular particles.
Analogous to realistic glass formers, it exhibits an initial rise due to individual particle swaps. It is followed by a plateau at intermediate time. The width of the plateau increases at lower temperatures. The plateau values of the MSD accounts for not only particles vibrating about their meta-stable positions but also a population of particles performing back-and-forth swaps among a small number of positions. It reaches the diffusive regime at long time, when most regular particles have been visited by some swap-initiators and have taken multiple swaps to escape any initial local traps. All these microscopic dynamics are observable from real-space visualizations of the regular particles.
The comparison of the MSD with and without swap is also shown in \fig{fig:msd_sisf}(a), from which we can observe that swaps dominate over MD dynamics.

    \Fig{fig:msd_sisf}(b) shows the self-intermediate scattering function defined as $F_s = \langle~\cos{(\textbf{k}\cdot[\textbf{r}_i(t)-\textbf{r}_i(0)])} ~\rangle$ averaged over regular particles $i$. Here, the magnitude of the wave vector $\textbf{k}$ is $k=2\pi/\lambda$ where $\lambda = 1$ is the length scale over which we measure each particle's overlap with its initial position. 
Similar to realistic glass formers, at low temperatures, it shows an initial decay, a plateau, and a final decay to zero at long time, which is related to the two-step rise in the MSD. The final decay of $F_s$ takes a stretched exponential form. 

\begin{figure}[tb]
	\includegraphics[width=\columnwidth]{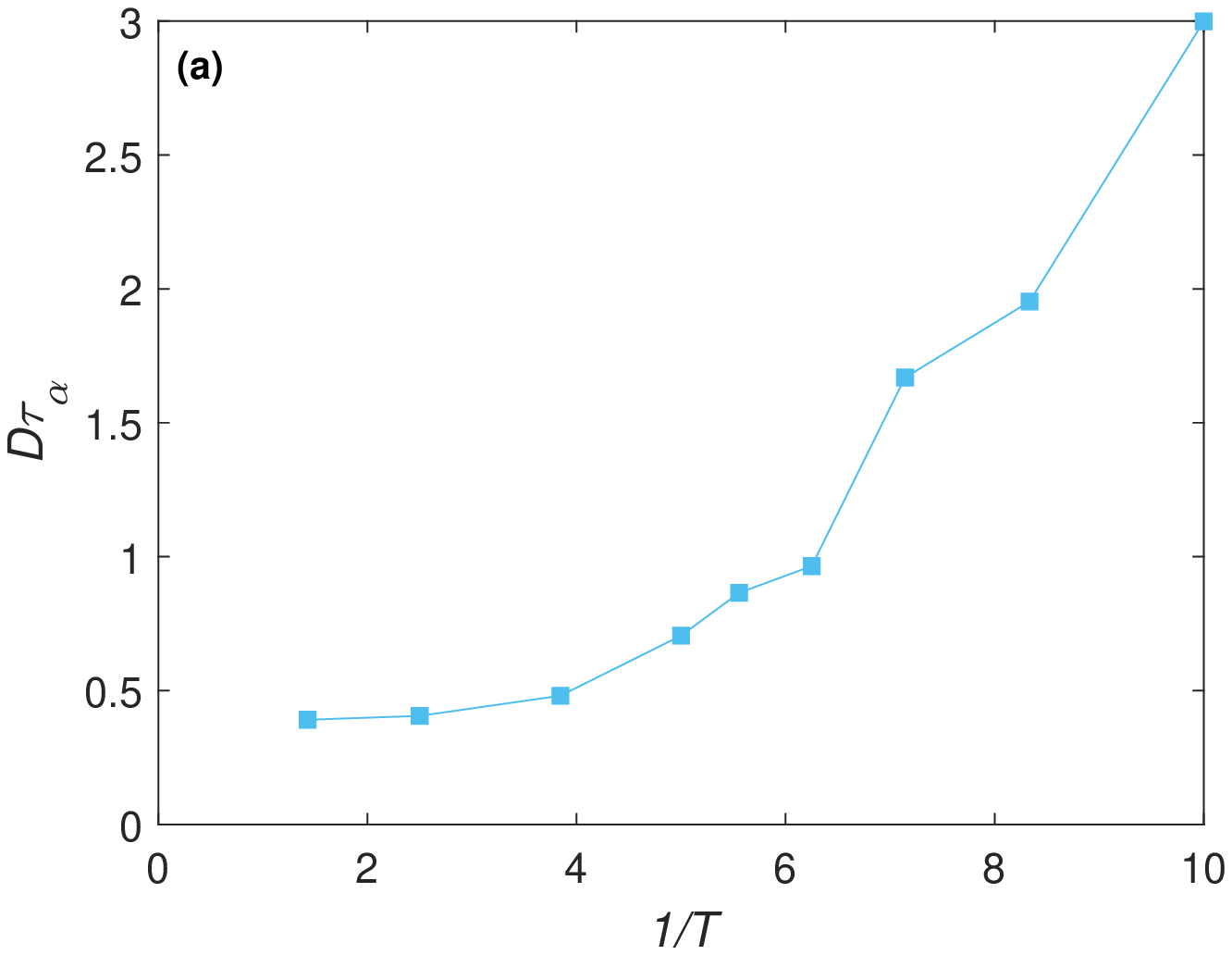}
	\includegraphics[width=\columnwidth]{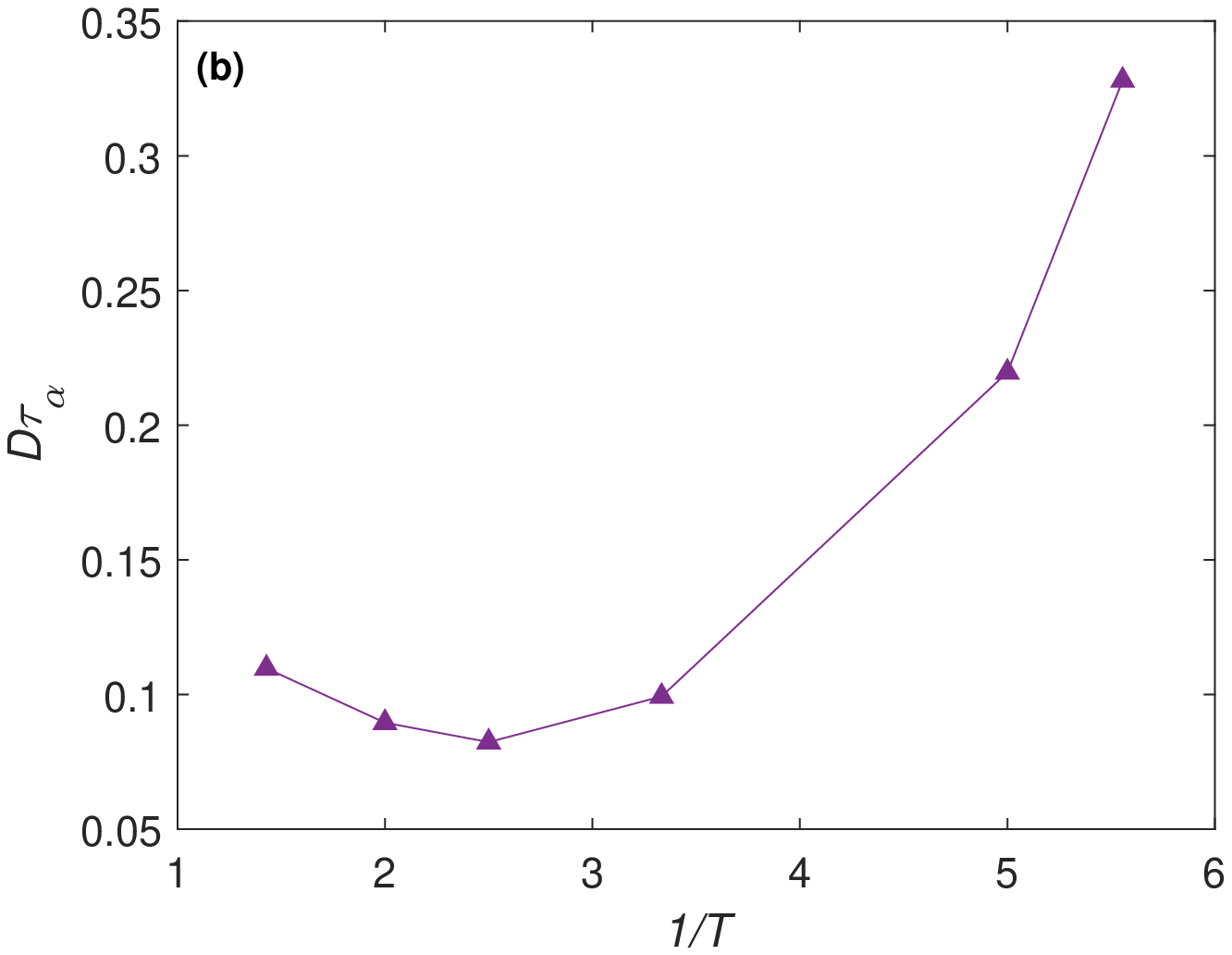}
	\caption{Stokes-Einstein violation in (a) the partial swap system with $\phi_s=0.017$ and (b) the MD system without swap, i.e. $\phi_s=0$.}
	\label{fig:SE}
\end{figure}

\begin{figure}[tb]
	\includegraphics[width=\columnwidth]{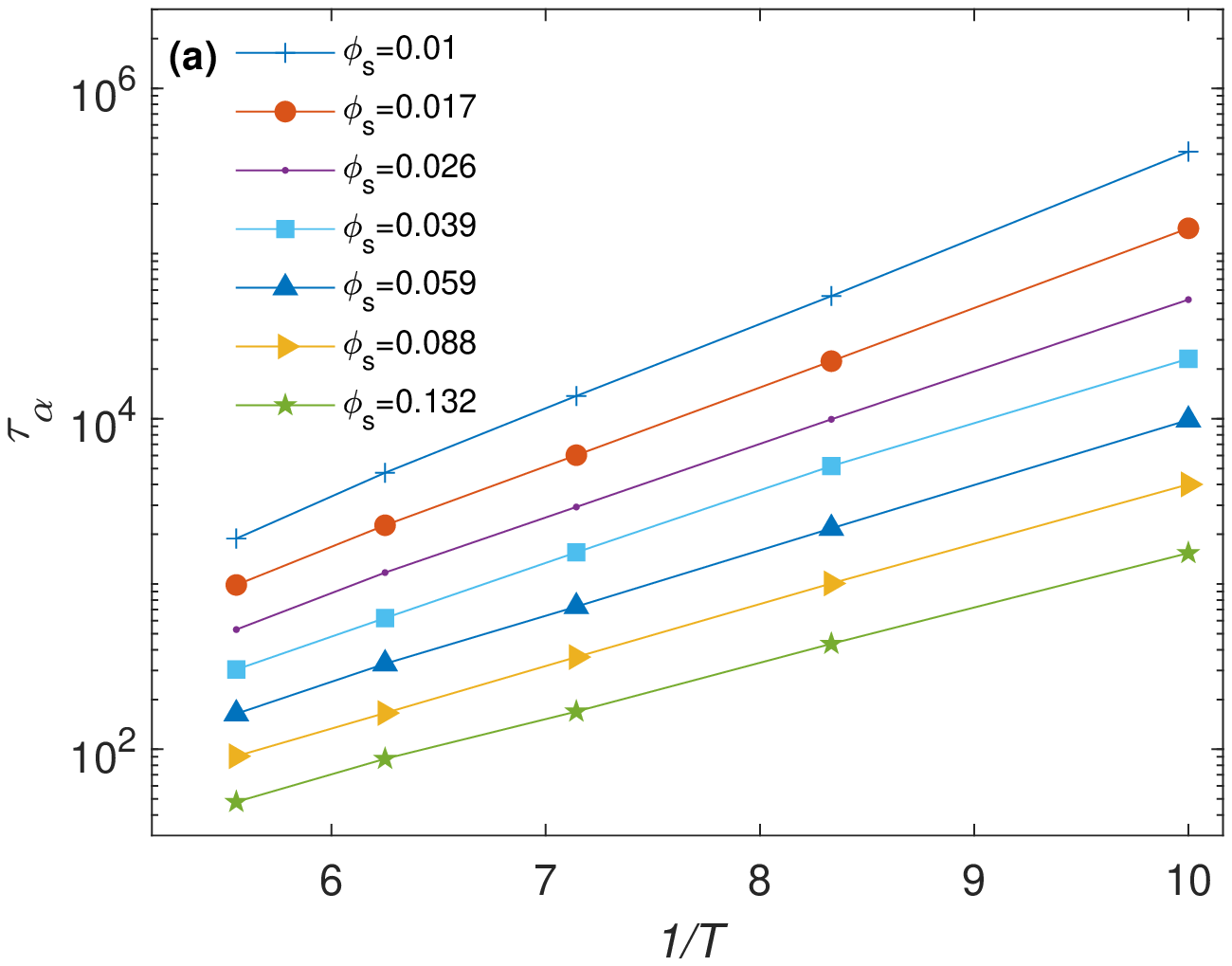}
	\includegraphics[width=\columnwidth]{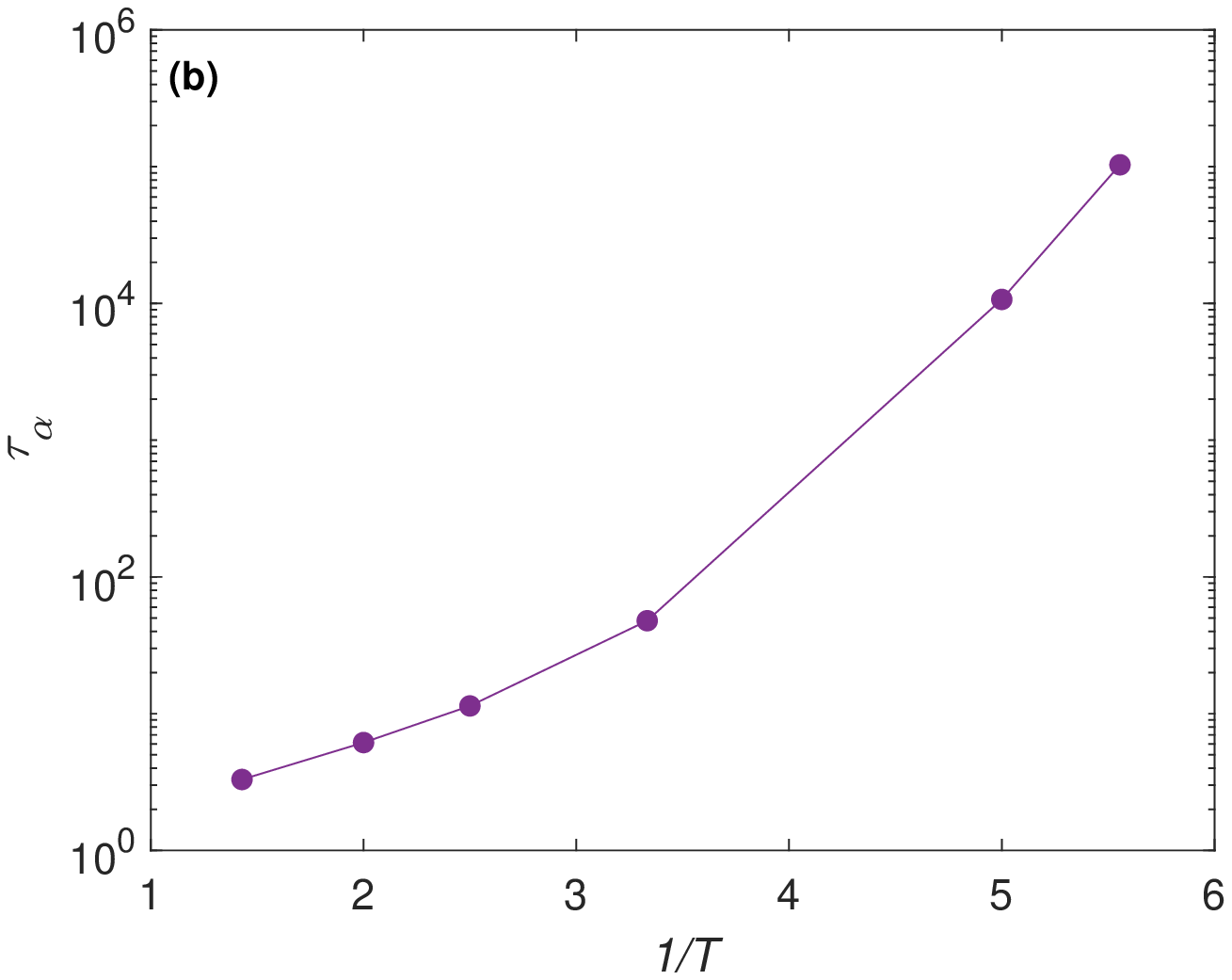}
	\caption{The structural relaxation time $\tau_{\alpha}$ as a function of temperature (a) with swap and (b) without swap.} 
	\label{fig:tau}
\end{figure}

\begin{figure}[tb]
	\centering
	\includegraphics[width=\columnwidth]{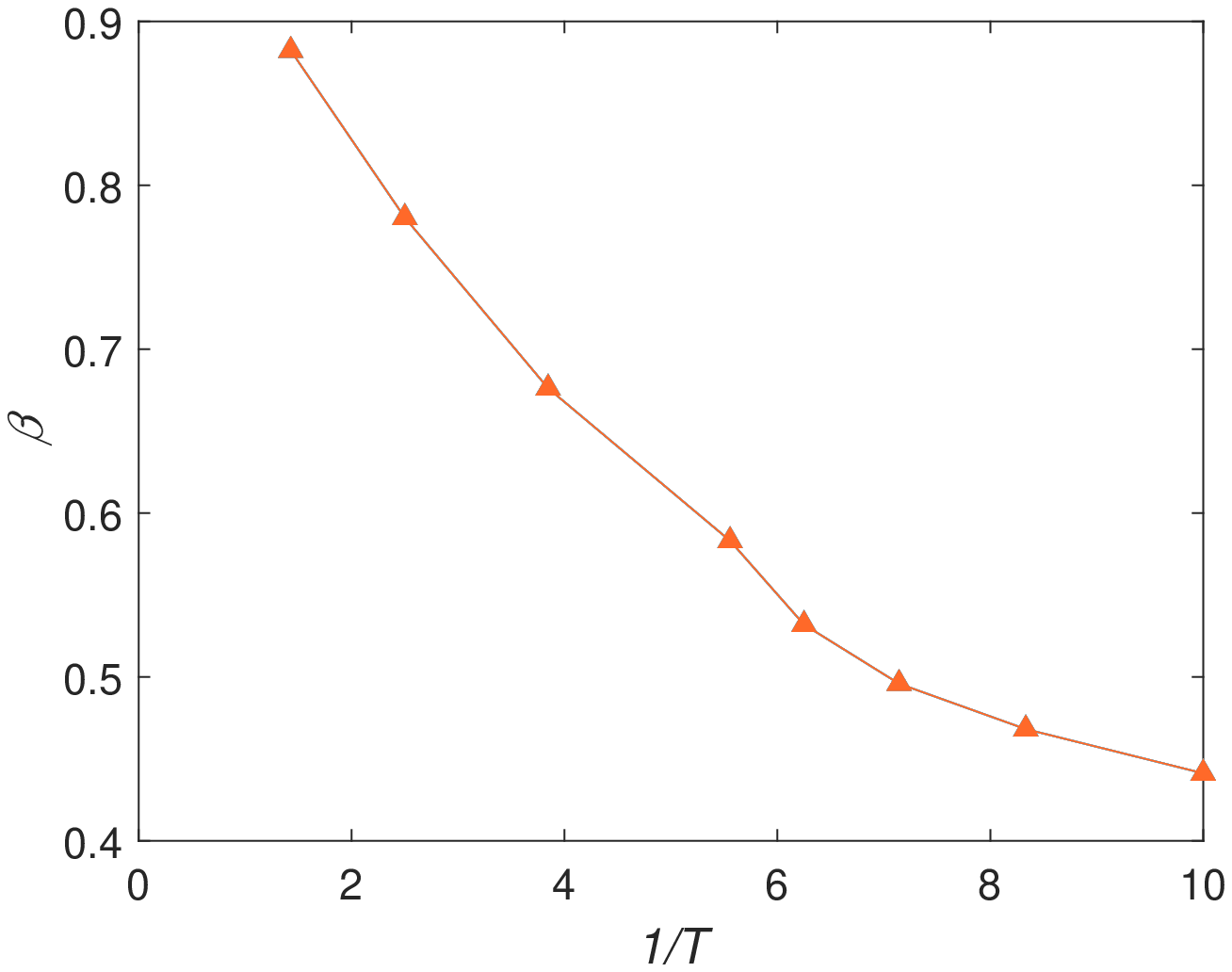}
	\caption{Stretching exponent $\beta$ as a function of temperature for $\phi_s=0.017$.}
	\label{fig:beta}
\end{figure}

 The partial swap system also shows a Stokes-Einstein violation, depicted in \ref{fig:SE}(a), showing the decoupling of local motions and long time structural relaxation. We have observed that it is a strong glass. This is illustrated in the straight lines in Figure \ref{fig:tau}(a), which shows a semi-log plot of the structural relaxation time $\tau_{\alpha}$ against inverse temperature. We obtain $\tau_{\alpha}$ by fitting the self-intermediate scattering function to the stretched exponential form $e^{-(t/\tau_{\alpha})^\beta}$. We use $\lambda=2$ for the wave vector $k=2\pi/\lambda$. In comparison, we plot $\tau_{\alpha}$ against 1/T for the case without swap. One can see that the presence of only MD motion produces a more fragile glass, as indicated by the curvature in \ref{fig:tau}(b). To further exemplify the glassiness of the system, we also plot the stretching exponent $\beta$ in Figure \ref{fig:beta}. As is typical of glasses, it becomes smaller at low T.


\begin{figure}[tb]
		\centering
		\includegraphics[width=0.8\columnwidth]{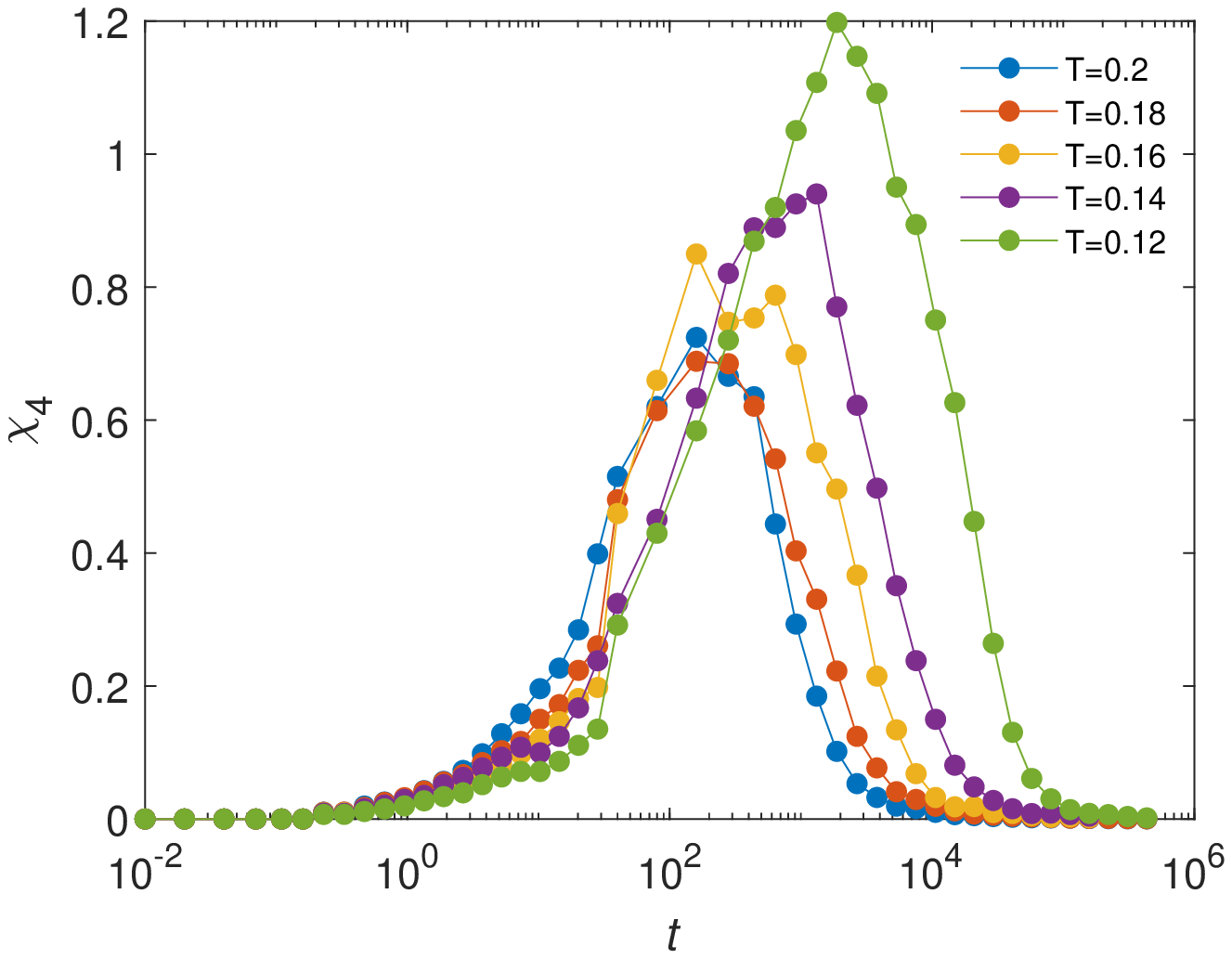}
		\caption{Four point susceptibility function $\chi_4$ at various temperature $T$ for $\phi_s$=0.059. The taller peaks at lower temperatures indicate increased dynamic heterogeneity.}
		\label{fig:chi4}
	\end{figure}

	Another archetypal feature of glass is dynamic heterogeneity. This is measured by the four-point susceptibility, which can be thought of as the magnitude of fluctuations in a particular overlap function. We define an overlap function $O(t)$ of particle positions as follows:
	\begin{equation}
		O(t) = \frac{1}{N}\Theta(|\textbf{r}_i(t)-\textbf{r}_i(0)|-r_{min})
	\end{equation}
	where $N$ is the total number of particles considered, $\Theta(r)$ is the Heaviside step function, and $r_{min}=1$ is a cutoff distance. By this definition, the overlap of an individual particle is 1 if the magnitude of its displacement is less than $r_{min}$,  and it is 0 once the particle exceeds this displacement. This form of the overlap function is akin to that used in Refs. \cite{narumi2011,shi2020}. The four-point susceptibility $\chi_4$ is then:
	\begin{equation}
		\chi_4 =  N\langle (O(t)-\overline{O}(t))^2 \rangle.
	\end{equation}
	Figure \ref{fig:chi4} shows the plot of $\chi_4$ for regular particles at various temperatures.
    We observe that the maximum height of $\chi_4$ increases as the temperature decreases. This is analogous to results from typical glass formers and indicates greater dynamic heterogeneity. These quantities obtained for the partial swap system fall within typical range of values for glass-forming models. We would like to mention that the glassy system formed by regular particles with partial swap may have properties, e.g. fragility, considerably different from the original non-swapping system, although they fall in the range of typical glasses.

	\begin{figure}[tb]
		\centering
		\includegraphics[width=\columnwidth]{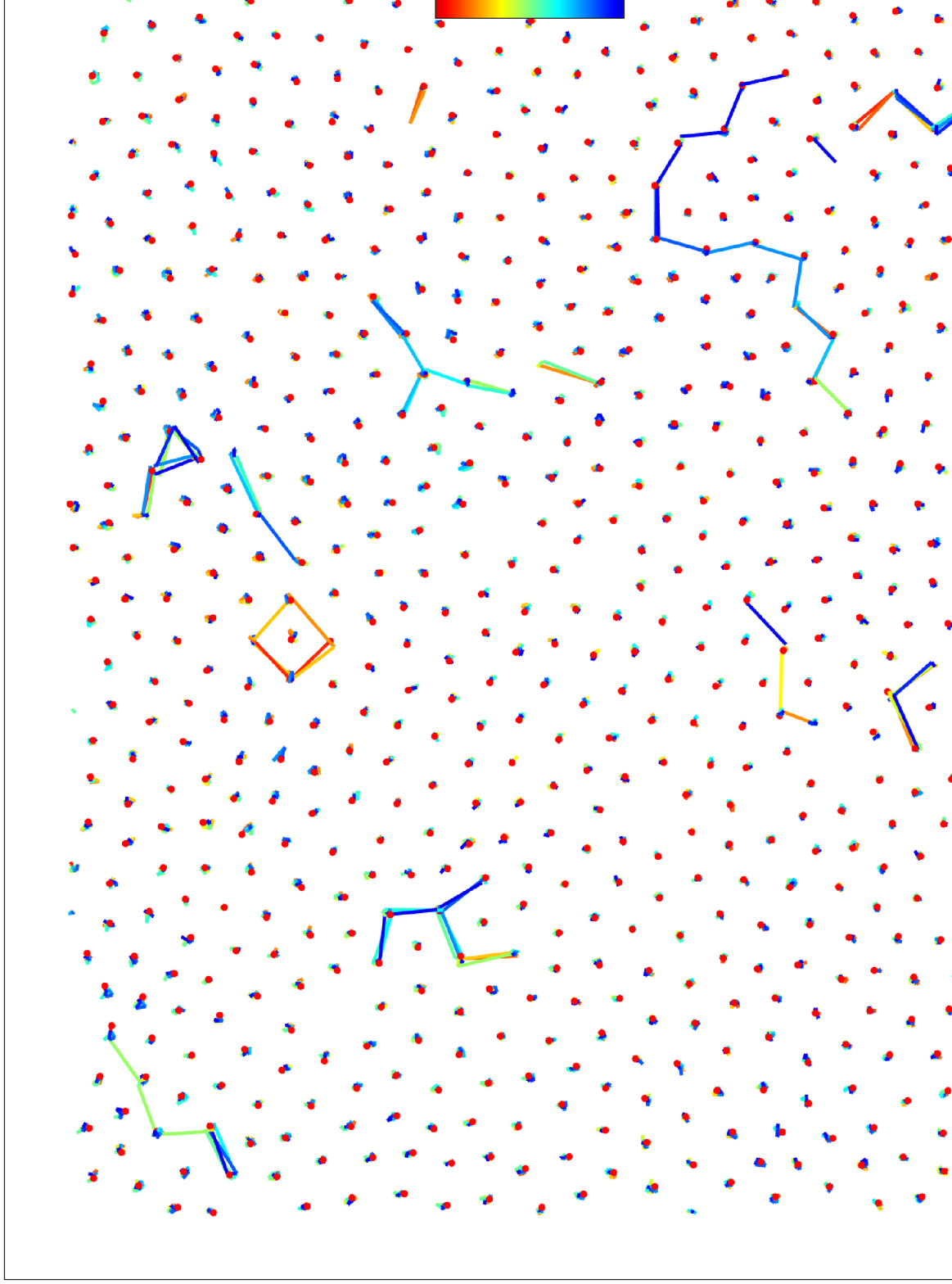}
		\caption{Time-colored trajectory plots for regular particles at $T=0.1$ and $\phi_s=0.026$ over a time window of $\Delta t=0.015$ showing string-like motions. Red dots show the initial positions of regular particles. The trajectory of each particle is illustrated with straight line segments joining a time sequence of particle positions, with each segment colored according to time. Most particles displace negligibly, but some show displacements comparable to particle separations. Displaced particles are often adjacent to each other, with trajectories well aligned forming strings. A long string created by 12 regular particles can be observed.}
		\label{fig:string}
	\end{figure}

    \newcommand{\rc}{\mathbf r^{c}}

	Our system of regular particles also exhibits string-like motions as already illustrated in the main text. \Fig{fig:string} further depicts them using a time-colored coarse-grained trajectory plot.
	To create the plot, we follow Refs. \cite{lam2017,yip2020}, with the trajectory of each particle drawn as straight line segments joining consecutive positions, each of which is colored according to time.
	In addition, red dots show the initial positions of all regular particles.
	Therefore, particles with small displacements simply show up as a red dot. Each particle taking part in a swap is represented by a red dot connected to a line segment of length comparable to the inter-particle separations, which are the typical swapping distances. The color of the dominant line segment indicates when the swap occurs.
	We observe in \fig{fig:string} trajectories in striking similarity with those in, e.g., glassy colloidal experiments \cite{yip2020}. In one case, individual trajectories of 12 particles join with nearly perfect alignment to form a long string. 

	In our partial-swap system, the cause of these string-like motions is easy to understand.
	For example, the longest string in \fig{fig:string} is created by a single initiator swapping one by one with 12 regular particles, the trajectory of each of which contributes to a segment in the string. A close examination also reveals some back-and-forth swapping during the course of movements of the swap-initiator, leading to some thickened trajectory lines.
	In some other cases, the strings have branched geometries. This arises when a swap-initiator displaces a line of particles, returns to one of its earlier positions, and then displaces another line of particles in a different direction. In contrast to displacement plots, trajectory plots show both the forward and backward trips and thus register  back-and-forth motions as well.

	\section{Comparison with Distinguishable-Particle Lattice Model}

	We have recently proposed a distinguishable particle lattice model (DPLM) \cite{zhang2017}, which exhibits a wide range of glassy phenomena, 
    including Kovacs paradox \cite{lulli2020}, Kovacs effect \cite{lulli2021}, a wide range of fragility \cite{lee2020}, heat capacity overshoot \cite{lee2021}, and low-temperature two-level systems \cite{gao2022}. Here, we generalize the DPLM to include local particle partial swap for a comparison with our MD results.
The purpose of this comparison is two fold. First, lattice models are more easily understandable in general and reproducing the MD results with the DPLM can be an important step for a thorough understanding of the MD results. Conversely, a successful lattice model should compare well with the widest possible range of phenomena, requiring few or no modification in the model definition. A successful comparison here can add to our previous efforts \cite{zhang2017,lulli2020,lulli2021,lee2020,lee2021,gao2022} and further supports the relevance of the DPLM to glass. 

	 
	 { For a comparison with MD simulations, we adopt  the DPLM defined in Ref. \cite{lee2020} for strong glass,  except that void-induced dynamics is now replaced by particle swaps. This alters the dynamics but not the thermodynamics. The modeled system is defined on a 2D square lattice with length $L=40$ and is fully occupied by $N= L^2$ particles without voids. The total energy of the system is given by:
	 	\begin{equation}
	 		E=\sum_{{<i,j>}'}V_{s_is_j}
	 	\end{equation} 
	 	where the sum is over occupied adjacent sites $i$ and $j$. The variable $s_i=1,2,\dots, N$ denotes which of the $N$ particles in the system is at site $i$. The  interaction energy $V_{kl}\in [0,1]$ between each particle pair $k$ and $l$ is sampled at the beginning of a simulation from a uniform distribution $g(V)$, leading to a strong glass \cite{lee2020}.
        
	\begin{figure}[tb]
		\centering
			\label{fig:power_law_regular_dplm}
			\includegraphics[width=\columnwidth]{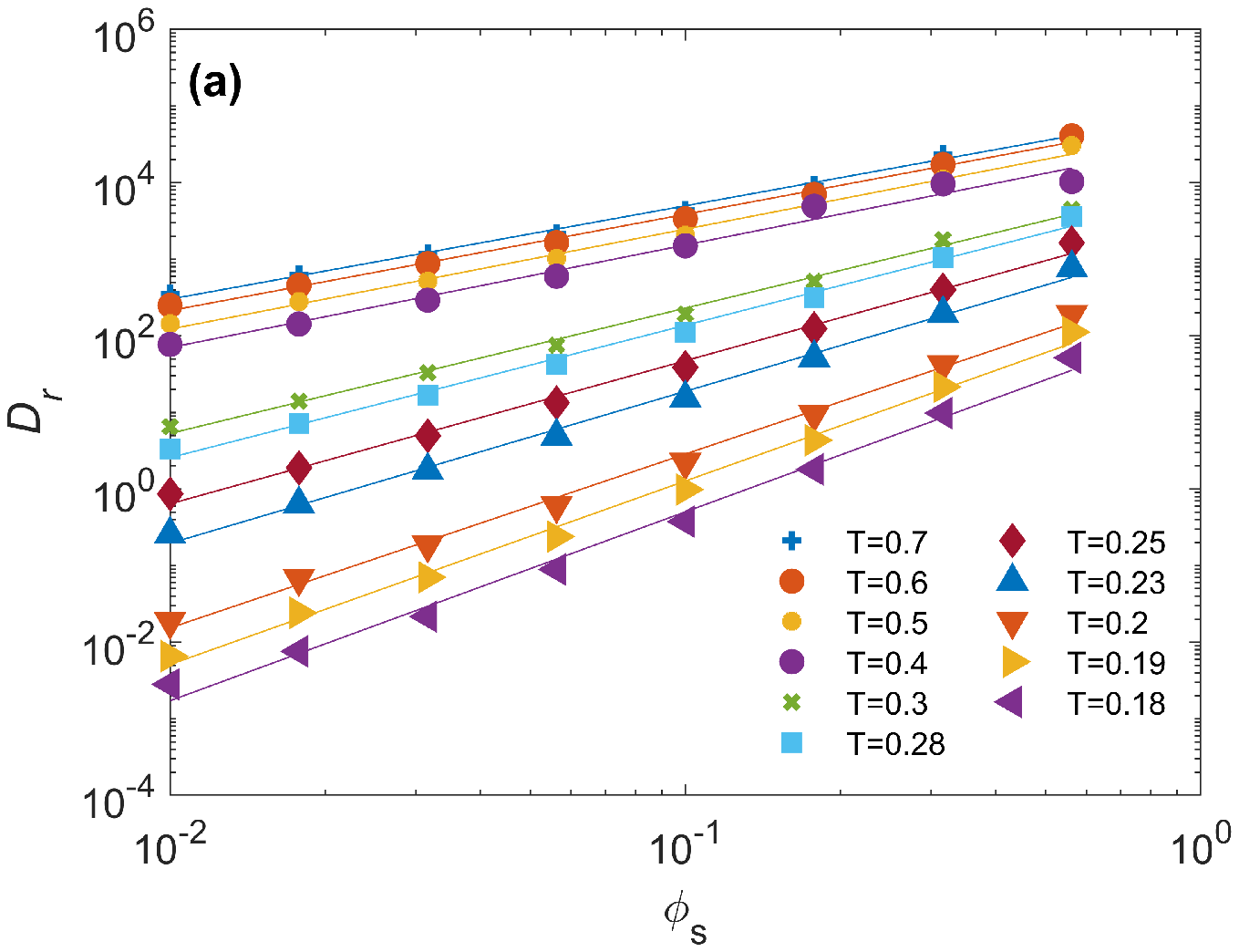}
			\label{fig:power_law_swapinit_dplm}
			\includegraphics[width=\columnwidth]{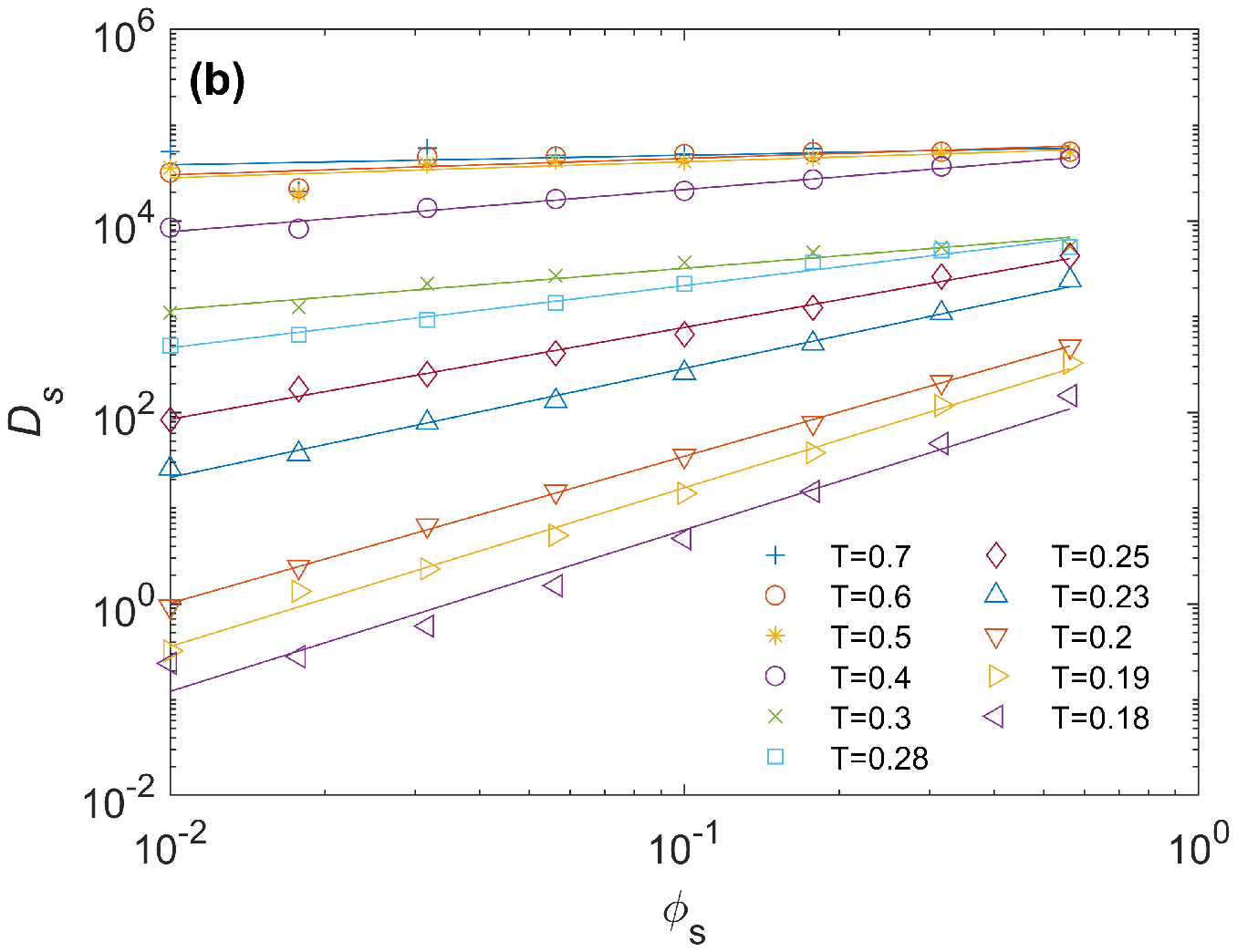}
		\caption{DPLM results on diffusion coefficients $D_r$ and $D_s$ of regular particles (a) and swap-initiators (b), showing power-law relations with respect to the fraction $\phi_s$ of swap-initiators.}
		\label{fig:power_law_dplm}
		
	\end{figure}

    To implement partial swap, a fraction $\phi_s$ of the $N$ particles are randomly pre-selected as swap-initiators. In the simulation, each Monte Carlo step corresponds to a time step $\Delta t=1/L^2w_0$, where $w_0=10^6$. At each time step, a pair of neighboring particles is chosen randomly. Similar to the algorithm in our MD simulations, if at least one particle in a pair is a swap-initiator, the particles are swapped with a probability $\mathrm{min} \left \{  1,e^{-\Delta E/ k_BT}\right \}$, where $\Delta E$ is the change of the system energy $E$ due to the swap. 
\begin{figure}[tb]
		\includegraphics[width=\columnwidth]{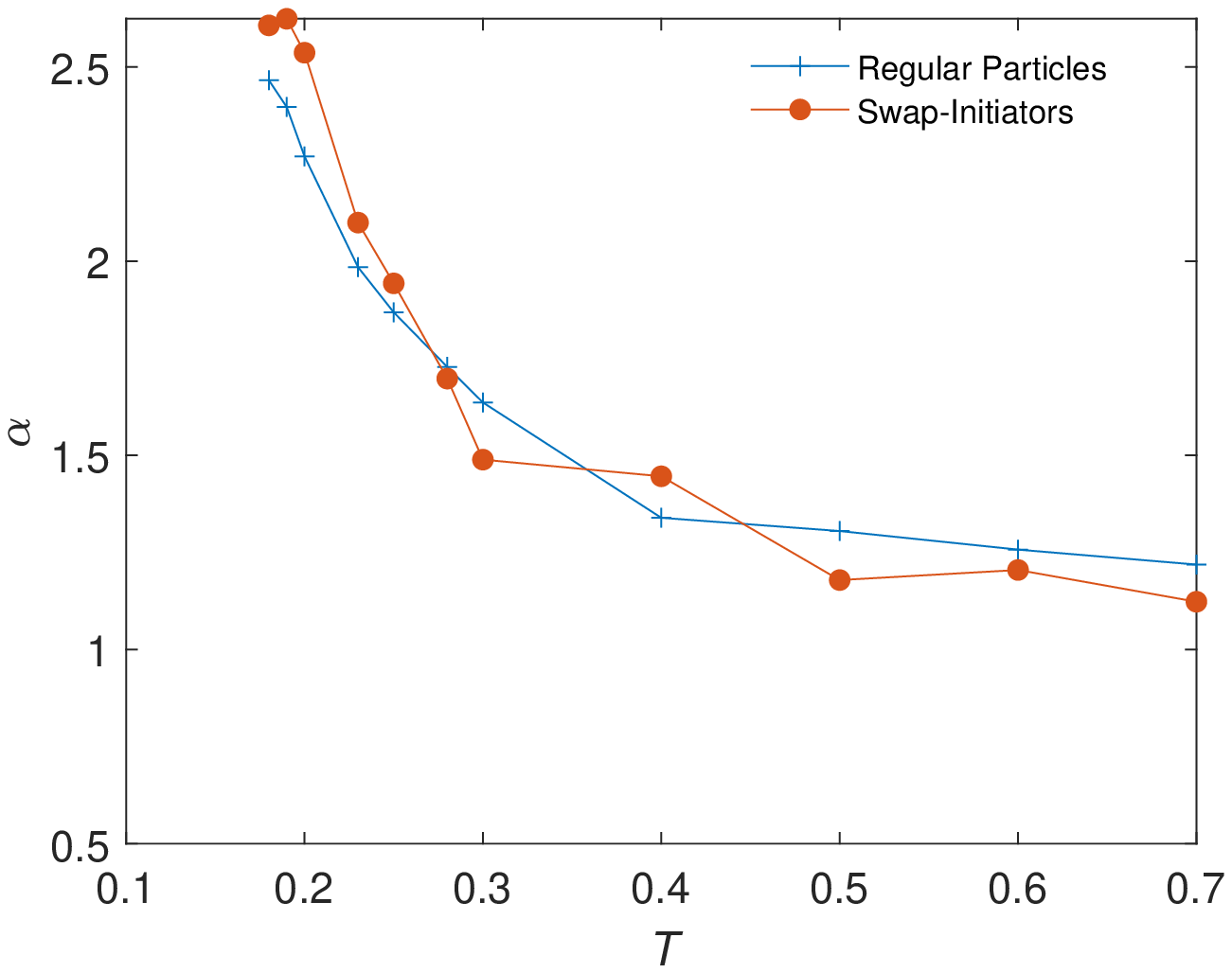}
		\caption{DPLM results on the power-law exponent $\alpha$ for regular particles (blue) and swap-initiators (orange) against temperature $T$.}
		\label{fig:alpha_dplm}
\end{figure}

We have measured the diffusion coefficients $D_r$ and $D_s$ of regular particles and swap-initiators and results are shown in Figure \ref{fig:power_law_dplm}. Similar to the MD results, we observe the power laws:
\begin{equation}
      D_r \sim \phi_{s}^{\alpha} , ~~~~~~~~~~~
		D_s \sim \phi_{s}^{\alpha-1}.
\label{Dr}
\end{equation}
The values of the exponent $\alpha$ obtained are shown in \fig{fig:alpha_dplm}, which  again shows strong resemblance to our MD results.


The DPLM with swap has thus successfully reproduced our main MD simulation results qualitatively as demonstrated above. There is however one notable difference: the facilitation in the DPLM is stronger than in MD simulations. At a sufficiently low $T$ corresponding to a scaling exponent $\alpha \simeq 2$, two nearby swap initiators in the DPLM almost always display much faster dynamics than isolated ones. This is closely analogous to the facilitation between two nearby voids, which has been illustrated in detail in Ref. \cite{zhang2017}. The strong facilitation stems from the very different energetic properties of the two particles. When a particle close to a swap-initiator is replaced by another due to the motion of a second initiator, the potential energy landscape experienced by the first initiator is usually changed  significantly in the DPLM. This opens up additional pathways of motions, leading to facilitation. In contrast,  there is much fluctuation in the facilitation in our MD simulations with swaps, as some pairs of nearby swap-initiators do not appear to facilitate the motions of each other. This is because only particles with similar radii are energetically capable to swap \cite{ninarello2017}. The motions of a swap-initiator thus only alter the configuration of the system slightly, as particles are replaced only by those with slightly different radii. The energy landscape thus changes little, resulting at a weaker facilitation. This may also explain a rather weak peak in $\chi_4$ in our MD simulations as observed above.

\bibliography{supplement}